%% file: main.tex
\documentclass[a4paper,10pt]{article}
\usepackage{cite}
\usepackage{amsmath,amssymb,amsfonts,amsthm}
\usepackage{graphicx}
\usepackage{textcomp}
\usepackage{hyperref}
\def\BibTeX{{\rm B\kern-.05em{\sc i\kern-.025em b}\kern-.08em
    T\kern-.1667em\lower.7ex\hbox{E}\kern-.125emX}}
\usepackage{algorithm}
\usepackage[noend]{algpseudocode}
\usepackage{flushend}
\usepackage{cuted}
\usepackage{mathtools}
\usepackage{widetext}
\usepackage{tikz}
\algblock{Input}{EndInput}
\algnotext{EndInput}
\algblock{Output}{EndOutput}
\algnotext{EndOutput}

\newtheorem{lemma}{Lemma}
\usepackage{bm}
\usepackage{braket}
\usepackage[version=3]{mhchem}
\usepackage{qcircuit}
\DeclareMathOperator{\tr}{tr}
\newcommand{\arrep}[1]{\ar @<4pt> @/^/[#1]|-{\mbox{$\times L$}}}
\newcommand{\lstickx}[1]{\lstick{\makebox[1.5em][l]{$#1$}}}
\newcommand{\UU}[1][]{\ensuremath{R_{\hat{\bm{n}}_{#1}}(\pi)}}
\newcommand{\Rn}[1][]{\ensuremath{R_{\hat{\bm{n}}_{#1}}(\theta_{#1})}}

\renewcommand{\baselinestretch}{1.0} 
\renewcommand{\title}[1]{\vspace{\fill}
\eject\addtolength{\baselineskip}{4pt}
{\bfseries\LARGE #1}\\[3mm]\addtolength{\baselineskip}{-4pt}}
\renewcommand{\author}[3]{\parbox[t]{75mm}
{\begin{center}{\scshape #1}\\[3mm] #2\\
 {\ttfamily #3} \end{center}}}

\begin{document}

\begin{center}
\title{Optimizing Parametrized Quantum Circuits \\
with Free-Axis Single-Qubit Gates}
\author{Hiroshi C. Watanabe$^*$}
{Quantum Computing Center, Keio University\\
JST PRESTO}
{hcwatanabe@keio.jp}
\author{Rudy Raymond$^*$}
{IBM Quantum, IBM Japan\\
Quantum Computing Center, Keio University\\
Dept. of Computer Science, The Univ. of Tokyo}
{rudyhar@jp.ibm.com}\\
\author{Yu-ya Ohnishi}
{Materials Informatics Initiative, RD Technology \& Digital Transformation Center, JSR Corporation, JSR BiRD}
{yuuya\_oonishi@jsr.co.jp}
\author{Eriko Kaminishi}
{Quantum Computing Center, Keio University\\
JST PRESTO}
{kaminishi@keio.jp}\\
\begin{center}
\author{Michihiko Sugawara}
{Quantum Computing Center, Keio University}
{sugawara.a6@keio.jp}
\end{center}
\end{center}

\def\thefootnote{*}\footnotetext{Equal contribution}
\def\thefootnote{\arabic{footnote}}
\begin{abstract}
Variational quantum algorithms, which utilize Parametrized Quantum Circuits (PQCs), are promising tools to achieve quantum advantage for optimization problems on near-term quantum devices. 
Their PQCs have been conventionally constructed from parametrized rotational angles of single-qubit gates around predetermined set of axes, and two-qubit entangling gates, such as CNOT gates. 
We propose a method to construct a PQC by continuous parametrization of both the angles and the axes of its single-qubit rotation gates.
The method is based on the observation that when rotational angles are fixed, optimal axes of rotations can be computed by solving a system of linear equations whose coefficients can be determined from the PQC with small computational overhead. 
The method can be further simplified to select axes freely from continuous parameters with rotational angles fixed to half rotation or $\pi$. 
We show the simplified free-axis selection method has better expressibility against other structural optimization methods when measured with Kullback-Leibler (KL) divergence.
We also demonstrate PQCs with free-axis selection are more effective to search the ground states of Hamiltonians for quantum chemistry and combinatorial optimization.
Because free-axis selection allows designing PQCs without specifying their single-qubit rotational axes, it may significantly improve the handiness of PQCs.
\end{abstract}


\section{Introduction}
\input{intro}

\section{Methods}\label{sec:methods}
\input{methods}

\section{Numerical Results}\label{sec:results}
\input{result_body}

\section{Conclusion}\label{sec:conclusion}
\input{conclusion}

\section*{Acknowledgment}
H.C.W. was supported by {JSPS}~under Grant Numbers~{20K03885} and {20H05518}, and~{JST PRESTO}~under Grant Number~{JPMJPR17GC}. 
E.K. was supported by {JSPS}~under Grant Number~{20K14388} and {JST PRESTO}~under Grant Number~{JPMJPR2011}. 
In addition, H.C.W., E.K., and M.S. were supported by the {MEXT Quantum Leap Flagship Program}~under Grant Number~{JPMXS0118067285} and {JPMXS0120319794}.
R.R. would like to thank colleagues at IBM Quantum, Yutaka Shikano, and Harumichi Nishimura for insightful technical discussion.
The authors would like to thank Naoki Yamamoto, Michael Lubasch, and Christiane Koch for their valuable comments. 

\section*{Code availability}
The Fraxis code used to create and the data in this article is executable on Python using Qiskit version 0.36~\cite{Qiskit}, and can be found on GitHub, 
\url{https://github.com/KQCC-Chemistry/SeQpt}.

\bibliographystyle{IEEEtran}
\bibliography{references.bib}

\end{document}

%% file: intro.tex
Parametrized quantum circuit (PQC) is one of the most essential components of hybrid quantum-classical algorithms on near-term quantum devices\cite{Cerezo2020arXiv, Bharti2021arXiv, TillyetalVQE2021}.
With PQC, a quantum state is expressed by a sequence of one- and two-qubit gates, in which the rotation angles and axes are classically controllable.
Variational quantum algorithms optimize these parameters so that their cost function is minimized.
For example, in variational quantum eigensolver (VQE)\cite{Peruzzo2014NatCom}, which treats fermionic or spin Hamiltonian of physics and chemistry, the cost function is the energy and the PQC represents its wavefunction.
PQC that gives the minimum energy corresponds to the wavefunction of the ground state according to the variational principle in VQE.
Quantum Approximate Optimization Algorithm (QAOA)\cite{QAOA2014} is another example using PQC.
Combinatorial optimization problems are targets of QAOA and the PQC generates a certain quantum state, for which the measurement yields a probability of each bit string encoding the solutions of the problems. %

The design of PQC is critical in variational quantum algorithms.
Oversimplified PQC cannot express the optimal quantum state even if it could be implemented on noisy quantum devices.
On the other hand, a PQC designed with a deep circuit for high expressibility cannot be implemented on currently-available noisy quantum devices.
A physics-based ansatz such as unitary coupled-cluster method in VQE is one of the examples of a deep circuit, in which the required number of two-qubit gates is too large to execute on near-term quantum devices\cite{Preskill2018quantumcomputingin}.
Highly expressible PQC can also be created in hardware efficient (or heuristic) ansatz\cite{Kandala2017Nat, Barkoutsos2018PRA, Ganzhorn2019PRApp, Gard2020npjQI, Tang2021PRXQ, Tkachenko2020arXiv} by applying multiple layers.
However, it often results in either too complicated landscape of cost function to find the global minimum, or barren plateau in which the gradient vanishes\cite{McClean2018NatComm, Cerezo2021NatComm, Pesah2020arXiv, Holmes2021PRL, Zhao2021arXiv, Cerezo2020arXiv, Arrasmith2020arXiv}.
Hardware-efficient ansatz also suffers from initial guess of parameters. Preparing a suitable set of 
initial parameters for typical sets of ansatz for various problem is challenging and has received a lot of attention. %

In recent years, various approaches are being explored to alleviate the problems in the design of PQCs.  
Adapt-VQE \cite{Grimsley2019NatComm} employs operator pool to achieve chemical accuracy with shallower circuit starting based on unitary coupled-cluster method.
It has been extended to hardware efficient ansatz~\cite{Tang2021PRXQ}.
Theoretical evaluations of expressibility of PQC is another direction to lead a proper design of PQC~\cite{Sim2019AQT, Funcke2021Quantum}.
The difficulty in optimizing the PQC with hardware efficient ansatz could be solved by sophisticating the optimization methods.
There are two main types of optimization methods that employ gradient-based or gradient-free methods.
For gradient-based optimization methods, analytical calculation of partial derivative in PQC\cite{Li2017PRL, Mitarai2018PRA} and stochastic gradient decent for VQE\cite{Sweke2020Quantum,Harrow2021PRL} are proposed and widely used.
Although naive gradient-based methods often suffers from barren plateau\cite{McClean2018NatComm, Cerezo2021NatComm, Pesah2020arXiv, Holmes2021PRL, Zhao2021arXiv, Cerezo2020arXiv, Arrasmith2020arXiv}, new approaches to avoid it, e.g., layerwise learning of quantum neural network\cite{Skolik2021QMI}, are still eagerly explored. Gradient-free optimizer, such as NFT\cite{Nakanishi2020PRR}, Rotosolve/select\cite{Ostaszewski2021Quantum}, and Jacobi-Anderson~\cite{Parrish2019arXiv}, is another option in classical optimization.
It is usually more robust against statistical error.

Nevertheless, while the design of PQCs has been optimized by using rotations of qubit states around predetermined axes~\cite{Nakanishi2020PRR,Ostaszewski2021Quantum}, the choice of such axes are arbitrary and the optimal ones in various problem settings, which are obviously influenced by the properties of the Hamiltonian, can be non trivial.
For example, the so-called $R_y$ (rotations around the y-axis) gates are popular for molecular Hamiltonians whose ground states are of real-valued probability amplitudes~\cite{Nakanishi2020PRR}, but other rotational gates (such as, $R_x$ and $R_z$) may be necessary for different types of Hamiltonians~\cite{Ostaszewski2021Quantum}.
Moreover, such additional rotations may facilitate shortcuts to quickly find the ground states as also argued in~\cite{Ostaszewski2021Quantum}. 

In this work\footnote{Preliminary results were presented in~\cite{WROKS2021}.}, we show it is possible to efficiently optimize over the continuous choices of rotational axes of single-qubit gates in a PQC and thus obtain a significant improvement over structural optimization using limited set of rotations in~\cite{Nakanishi2020PRR, Parrish2019arXiv, Ostaszewski2021Quantum}.
The proposed method, which we refer to as \textit{Free-Axis Selection} (or, in the hereafter shortened to \textbf{Fraxis}), handles the task to find an optimal axis of rotation with a given rotational angle $\theta$ on a single-qubit gate of a PQC.
It turns out that an optimal axis of rotation $\hat{\bm{n}} \in \mathbb{R}^3$ satisfies a linear system of equations ${\bm{R}}(\theta)\hat{\bm{n}} = {\bm{b}}(\theta)$, where the elements of the matrix ${\bm{R}}(\theta)$ and the vector ${\bm{b}}(\theta)$ can be determined by running the slightly-modified PQC.
Furthermore, the optimal axis is an eigenvector of ${\bm{R}}(\pi)$ when the angle $\theta$ is limited to one-half of full rotation.
Surprisingly, even under such limitaton, we show that PQCs with Fraxis have better expressibility than those with other structural optimization, and can find ground states relatively faster for several important Hamiltonians.

The benefits of PQC with Fraxis will be undoubtedly optimal when the underlying quantum devices natively support free-axis single-qubit gates. However, PQC with Fraxis still brings some important benefits even without such native implementability. Firstly, it frees users from specifying which set of rotational gates to choose from and thus lowers the barrier in designing PQCs on near-term quantum devices. Secondly, even though Fraxis gates are eventually decomposed into fixed-axis single-qubit gates on available quantum devices (we will show the results of running PQCs with Fraxis gates on IBM Quantum devices), our results imply that the optimization of their parameters with Fraxis is more efficient than corresponding techniques on fixed-axis single-qubit gates. This is not only because Fraxis has more degrees of freedom, but also because it considers correlations between parameters in the optimization. The decomposition of Fraxis gates may result in additional single-qubit gates, but because of the high quality of single-qubit gates on available quantum devices, this is not a problem. On the contrary, Fraxis improves the expressibility of PQCs without the need of adding layers of noisy two-qubit entangling gates. Thirdly, Fraxis gives a new formulation of sequential optimization for PQCs in the framework of matrix factorization that has opened paths for optimizing full degrees of freedom in single-qubit gates for quantum time simulation~\cite{WROKSYW2022}, and PQCs~\cite{WRSW2022}, albeit with more circuit evaluations.

The rest of the paper is arranged as follows. In Sec.~\ref{sec:methods}, we describe theoretical foundation of our proposed methods by first giving motivating examples, and then deriving a system of linear equations that makes free-axis selection possible in an efficient manner. We then show the simplified version of Fraxis whose computational overhead is slightly better than known methods. In Sec.~\ref{sec:results}, we demonstrate numerical experiments showing the power of the proposed methods in comparison to known ones. Namely, Fraxis have better expressibility and they can be used to efficiently search for ground states of Hamiltonians for quantum chemistry and combinatorial optimization. Finally, the paper is concluded in Sec.~\ref{sec:conclusion} in which we also discuss possible future work extending the proposed methods.  

%% file: methods.tex
We consider the problem of optimizing a PQC that  consists of a sequence of parametrized unitary gates $U = U_1 \ldots U_D$ 
so that from a given initial state $\rho$ the PQC generates a quantum state $U \rho U^\dag$ minimizing an objective function encoded in a Hamiltonian $M$. 
Namely, the value of $\left< M\right> \equiv \tr(M U\rho U^\dag)$ is minimized over all parameter space of the PQC.
We first present simple examples on why our proposed Fraxis can lead to more efficient PQCs and describe the basics of optimizing PQCs by borrowing the techniques in \cite{Ostaszewski2021Quantum}.
We then derive Fraxis in its general form, and present its simplified form that turns out to be computationally more efficient than the Rotoselect \cite{Ostaszewski2021Quantum}. 

\subsection{Motivating Examples}
\begin{figure}[tb]
  \centering
  \begin{tabular}{cc}
   \includegraphics[scale=0.5]{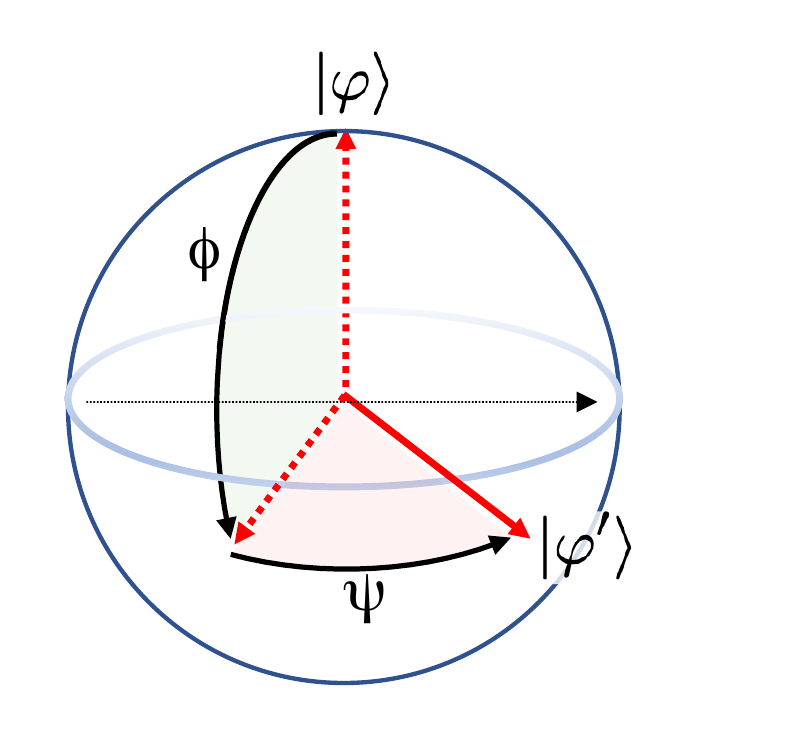} & 
   \includegraphics[scale=0.5]{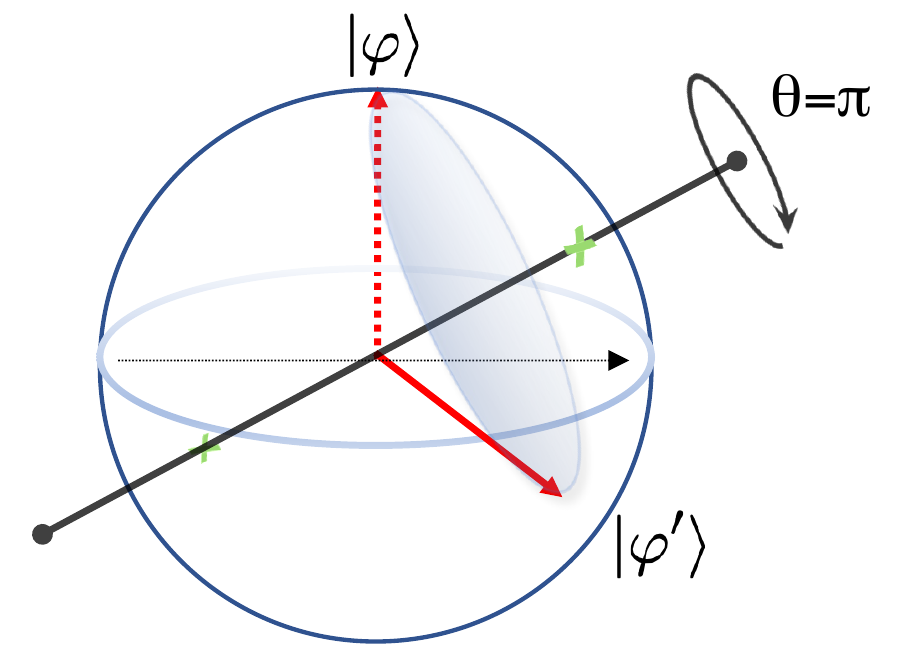}\\
   (a) Ry-Rz rotations &
   (b) Fraxis rotation
  \end{tabular}
  \caption{A comparison of obtaining a quantum state $\ket{\varphi'}$ from (a) rotating $\ket{\varphi}$ along the y-axis and then along the z-axis, i.e., applying a sequence of single-qubit gates $R_y(\phi)$ and $R_z(\psi)$ , and from (b) rotating $\ket{\varphi}$ with a Fraxis gate.}\label{fgr:ryrz-fraxis}
\end{figure}

Here is an example why selecting axes of rotations can be better than fixing the axes and optimizing the continuous angles illustrated in Fig.~\ref{fgr:ryrz-fraxis}.
Let us consider a simple toy Hamiltonian $H = X + Y + Z$ of a single qubit system.
The ground state of $H$ can be computed from a single-qubit PQC with $R_yR_z$ ansatz. 
Namely, a typical PQC consists of a $R_y(\theta_1)$ gate followed by a $R_z(\theta_2)$ gate.
Notice that if the initial quantum state is $\left|0\right>$, then clearly no PQCs with single $R_y$ or $R_z$ gate can find the ground state $\rho = \frac{1}{2}\left(I - \frac{1}{\sqrt{3}}\left(X + Y +Z\right)\right)$. 

This can be seen from the Bloch-vector representation of the ground state, which is $\bm{r}_{g} = -\frac{1}{\sqrt{3}}(1, 1, 1)^T$, and that of the initial state, which is $\bm{r}_{0} = (0, 0, 1)^T$.
The $R_y$ and $R_z$ gates rotate a quantum state along the $y$ and $z$ axes, respectively.
Thus, it is not possible to obtain $\bm{r}_{g}$ from $\bm{r}_0$ by either $R_y$ or $R_z$ alone (or, even by applying $R_z$ and $R_y$ gate in that order).
Thus, previously known approaches, such as, the Rotosolve and Rotoselect~\cite{Ostaszewski2021Quantum}, need both  $R_y$ and $R_z$ rotational gates, must order them correctly, and tune the angles to obtain the ground state. See the left figure of Fig.~\ref{fgr:ryrz-fraxis}. 

On the other hand, the effect of two rotations is the same as that of a single rotation due to the consequence of Euler's rotation theorem. It is well known that single-qubit unitaries are rotations of the Bloch sphere~\cite{nielsen_chuang_2010}, and each rotation can be represented by a real unit vector $\hat{\bm{n}} = (n_x, n_y, n_z)^T \in \mathbb{R}^3$ and an angle $\theta \in [0,2\pi]$ so that the unitary is
\begin{equation}\label{eqn:general-rotation}
    R_{\hat{\bm{n}}}(\theta) = \cos(\theta/2)I - i\sin(\theta/2)\left(n_x X + n_y Y + n_z Z \right). 
\end{equation}
Notice that the $R_y(\theta)$ and $R_z(\theta)$ are instances of the above unitary.
$R_{\hat{\bm{n}}}(\theta)$ rotates its input state by $\theta$ about the $\hat{\bm{n}}$ axis.
With regards to the toy Hamiltonian $H$, by some algebra we can confirm that rotating the input state $\left|0\right>$ by $\theta=\pi$ about the axis $\hat{\bm{n}} = \left(\frac{1}{\sqrt{2}}\sqrt{\frac{1}{2}+\frac{1}{2\sqrt{3}} }, \frac{1}{\sqrt{2}}\sqrt{\frac{1}{2}+\frac{1}{2\sqrt{3}} }, -\sqrt{\frac{1}{2} - \frac{1}{2\sqrt{3}}}\right)^T$ results in the ground state. See the right figure of Fig.~\ref{fgr:ryrz-fraxis}. 
Therefore, if we can optimize the axis of rotation of the corresponding unitary then we can obtain the ground state using only a single gate. 
Even if the quantum devices do not natively support free-axis gates, we can still obtain benefits in the formulation of circuit optimization with Fraxis.
We will show and argue later that optimizing the axis of rotation indeed results in quantum circuits with better expressibility and convergence.
In what follows, we first describe the steps to select axis of rotation in PQCs. 

\subsection{Optimizing PQCs}
The PQCs are used to find an $n$-qubit quantum state $\rho \in \mathbb{C}^{2^n\times 2^n}$ to minimize an objective function encoded in a Hermitian matrix $M$.
We follow a similar setting as the one in~\cite{Ostaszewski2021Quantum}.
The quantum state $\rho$ is generated by a PQC that consists of a sequence of unitary gates $U = U_1 U_2 \ldots U_D$ acting on $n$ qubits so that each $U_d$ is either a fixed two-qubit gate, or a parametrized single-qubit gate. Here we make a simplification for ease of reading as $U_d$ is in fact a $2^n \times 2^n$ matrix which is a tensor product of two-qubit gate or a single-qubit gate acting on a particular qubit with $2\times 2$ identity matrix acting on other qubits.
For ease of explanation, we consider only the single-qubit gates and consider the $d$-th unitary $U_d$ as a rotational gate with parameter $\theta_d$ written as:
\setlength{\arraycolsep}{0.0em}
\begin{eqnarray*}
   U_d \equiv e^{-i\frac{\theta_d}{2} \hat{\bm{n}}_d \cdot \overrightarrow{\sigma}} 
    &=&~\cos{\left({\theta_d/2}\right)}I - i \sin{\left({\theta_d/2}\right)} \hat{\bm{n}}_d \cdot \overrightarrow{\sigma}\nonumber\\
    &=&~R_{\hat{\bm{n}}_d}(\theta_d),
\end{eqnarray*}
\setlength{\arraycolsep}{5pt}
where $\hat{\bm{n}}_d \cdot \overrightarrow{\sigma} = n_{d,x} X + n_{d,y} Y + n_{d,z} Z$, for $\hat{\bm{n}}_d \in \mathbb{R}^3$ so that $|\hat{\bm{n}}_d|=1$.   

We want to optimize the following expected value on the objective function over the choice of parameters of all single qubit gates, i.e., $\theta_d$ and $\hat{\bm{n}}_d$ for $d=1,\ldots, D$, that we can gather as $\bm{\theta} = \left(\theta_1, \ldots, \theta_D\right)$ and $\left(\hat{\bm{n}}_1, \ldots, \hat{\bm{n}}_D\right)$.
To further simplify the notation, similar to~\cite{Ostaszewski2021Quantum} we write the expectation value as 
\setlength{\arraycolsep}{0.0em}
\begin{eqnarray*}
    \left<{M}\right> ~&=&~ 
    \mbox{tr} \left( {M}U_D\ldots U_{d+1}U_dU_{d-1}\ldots U_1 {\rho} U_1^\dagger\ldots U_{d-1}^\dagger U_d^\dagger U_{d+1}^\dagger \ldots U_D^{\dagger} \right),
\end{eqnarray*}
\setlength{\arraycolsep}{5pt}
which can be transformed as below because trace is invariant under cyclic permutations. 
\begin{equation*}
    \braket{M} = \braket{\hat{M}} = \mbox{tr}\left(\hat{M}U_d \hat{\rho} U_d^\dagger \right).
\end{equation*}
In the above, we utilize the following definitions
\setlength{\arraycolsep}{0.0em}
\begin{eqnarray*}
    \hat{M} &{}\equiv{}& U_{d+1}^\dagger \ldots U_D^\dagger {M} U_D \ldots U_{d+1},\\
    \hat{\rho} &{}\equiv{}& U_{d-1}\ldots U_1 {\rho} U_1^\dagger \ldots U_{d-1}^\dagger.
\end{eqnarray*}
\setlength{\arraycolsep}{5pt}
In the hereafter, we simply write $\hat{M}$ and $\hat{\rho}$ as $M$ and $\rho$, respectively, as they are clear from the context. 
Fixing all parameters of single-qubit gates excepting the $d$-th ones, the value $\left<M\right>$ is a function of $\theta_d$ and $\hat{\bm{n}}_d$ as below (see~\cite{Ostaszewski2021Quantum} for the proof).   
\begin{eqnarray}
    \left<M\right>_{\hat{\bm{n}}_d,\theta_d} &=&  \cos^2{\left(\theta_d/2\right)}~\mbox{tr}\left(M\rho\right)  + i\sin{\left(\theta_d/2\right)}\cos{\left(\theta_d/2\right)}\mbox{tr}\left(M\left[\rho, H_d\right] \right) + \sin^2{\left(\theta_d/2\right)}\mbox{tr}\left(MH_d\rho H_d\right),\label{eqn:obj-H-theta}
\end{eqnarray}
where $H_d \equiv \hat{\bm{n}}_d \cdot \overrightarrow{\sigma}$, and $\left[A, B\right] \equiv A B - B A$. 

Now, we remark an important result of~\cite{Ostaszewski2021Quantum,Nakanishi2020PRR}: given a fixed $\hat{\bm{n}}_d$ the optimal $\theta^*_d$ can be computed efficiently because $\left<M\right>_{\theta_d}$ is a sinusoidal form that can be fully characterized by evaluating the circuits on three different parameters of $\theta_d$.
The above fact is used to derive structural optimization of PQCs for finding the optimal angles $\bm{\theta}$ under fixed axes of rotations (e.g., Rotosolve in~\cite{Ostaszewski2021Quantum}), and for choosing $\left(\hat{\bm{n}}_1, \ldots, \hat{\bm{n}}_D\right) \in \{R_x, R_y, R_z\}^D$ where each of $R_x, R_y$ and $R_z$ corresponds to rotational axes $\hat{\bm{n}}^{(x)} = (1, 0, 0)^T$, $\hat{\bm{n}}^{(y)} = (0, 1, 0)^T$, and $\hat{\bm{n}}^{(z)} = (0, 0, 1)^T$, respectively.
In the next section we show how we can further extend the choices of rotational axes from all unit vectors $\hat{\bm{n}}_d \in \mathbb{R}^3$. 

\subsection{The Proposed Method: Free-Axis Selection}
Hereafter we describe how to optimize over the choice of a unit vector $\hat{\bm{n}}_d \in \mathbb{R}^3$ that defines the rotation axis with regards to a fixed rotation angle $\theta_d$.
For simplicity, we drop the subscript $d$. We want to determine the values of $n_x, n_y$ and $n_z$ to minimize $\left<M\right>_{\hat{\bm{n}}}$. 

By the Lagrange multipliers, the objective function to optimize is 
\setlength{\arraycolsep}{0.0em}
\begin{equation}
    \left(\hat{\bm{n}}^*, \lambda^*\right) = \min_{\hat{\bm{n}}, \lambda} f(\hat{\bm{n}}, \lambda)
    \equiv{} \left<M\right>_{\hat{\bm{n}}} - \lambda\left( n_x^2 + n_y^2 + n_z^2 - 1\right),
\end{equation}
\setlength{\arraycolsep}{5pt}
where $\lambda$ is a scalar whose value can be determined later. 

For ease of notation, let us define $\alpha_\theta \equiv i \sin{\left(\frac{\theta}{2}\right)} \cos{\left(\frac{\theta}{2}\right)}$, and assume that $\alpha_{\theta} \neq 0$ (it is zero if $\theta = 0$, where there is no need to change axis of rotation, or if $\theta = \pi$ which we will deal later).
Differentiating $f$ with respect to $n_x$, $n_y$, and $n_z$ we obtain 

\begin{widetext}
\begin{multline}\label{eq:dfdx}
    \frac{\partial f}{\partial n_x} = \alpha_\theta \mbox{tr}\left(M\left[\rho, X\right] \right) \\+ \sin^2{\left(\frac{\theta}{2}\right)}\left(2n_x \mbox{tr}\left( MX\rho X\right) + n_y \mbox{tr}\left(MX\rho Y + MY\rho X \right) + n_z\mbox{tr}\left( MX\rho Z + MZ\rho X\right)\right)
    - 2\lambda n_x,
\end{multline}
\begin{multline}\label{eq:dfdy}
    \frac{\partial f}{\partial n_y} = \alpha_\theta \mbox{tr}\left(M\left[\rho, Y\right] \right) \\+ \sin^2{\left(\frac{\theta}{2}\right)}\left(2n_y \mbox{tr}\left( MY\rho Y\right) + n_x \mbox{tr}\left(MX\rho Y + MY\rho X \right) + n_z\mbox{tr}\left( MY\rho Z + MZ\rho Y\right)\right) - 2\lambda n_y,
\end{multline}
\begin{multline}\label{eq:dfdz}
    \frac{\partial f}{\partial n_z} = \alpha_\theta \mbox{tr}\left(M\left[\rho, Z\right] \right) \\
    + \sin^2{\left(\frac{\theta}{2}\right)}\left(2n_z \mbox{tr}\left( MZ\rho Z\right) + n_x \mbox{tr}\left(MX\rho Z + MZ\rho X \right) + n_y\mbox{tr}\left( MY\rho Z + MZ\rho Y\right)\right) 
    - 2 \lambda n_z.
\end{multline}
\end{widetext}

At optimality, the value $\hat{\bm{n}}^* = (n_x^*, n_y^*, n_z^*)^T$ satisfies:
\begin{equation}\label{eq:partial-zero}
    \frac{\partial f}{\partial n_x} = \frac{\partial f}{\partial n_y} = \frac{\partial f}{\partial n_z} = 0.
\end{equation}

Defining $r_x, r_y, r_z, r_{(x+y)}, r_{(x+z)}$ and $r_{(y+z)}$ as in~\eqref{eq:relement_start}--\eqref{eq:relement_stop}, and noticing the following identities: 
\begin{eqnarray*}
    \mbox{tr}\left( MX\rho Y+ MY\rho X\right) &=& 2r_{(x+y)} - r_x - r_y,\\
    \mbox{tr}\left( MX\rho Z+ MZ\rho X\right) &=& 2r_{(x+z)} - r_x - r_z,\\
    \mbox{tr}\left( MY\rho Z+ MZ\rho Y\right) &=& 2r_{(y+z)} - r_y - r_z,
\end{eqnarray*}
we can arrange~\eqref{eq:dfdx}--\eqref{eq:dfdz} and~\eqref{eq:partial-zero} to obtain the following equation.  

\begin{widetext}
\begin{equation}\label{eqn:lin-axis}
\left[
    \sin^2{\left(\frac{\theta}{2}\right)}
    \begin{pmatrix}
    2r_x  & 2r_{(x+y)} - r_x - r_y & 2r_{(x+z)} - r_x - r_z\\
    2r_{(x+y)} - r_x - r_y & 2r_y & 2r_{(y+z)} - r_y - r_z \\
    2r_{(x+z)} - r_x - r_z & 2r_{(y+z)} - r_y - r_z & 2r_z
    \end{pmatrix}
    - 2\lambda^* \mathbf{I}
\right]
    \begin{pmatrix}
    n_x^*\\
    n_y^*\\
    n_z^*
    \end{pmatrix} = 
    -\alpha_\theta 
    \begin{pmatrix}
    \mbox{tr}\left(M\left[\rho, X\right]\right)\\
    \mbox{tr}\left(M\left[\rho, Y\right]\right)\\
    \mbox{tr}\left(M\left[\rho, Z\right]\right)\\
    \end{pmatrix}
\end{equation}
\end{widetext}

In~\eqref{eqn:lin-axis}, we use the following definitions which are the expectation values obtained from running the circuits each of which its $d$-th single-qubit gate is replaced with $X, Y, Z, (X+Y)/\sqrt{2}, (X+Z)/\sqrt{2}$ and $(Y+Z)/\sqrt{2}$ gate, respectively. 

\begin{eqnarray}
    r_x &\equiv& \mbox{tr}\left(MX\rho X\right),\label{eq:relement_start}\\
    r_y &\equiv& \mbox{tr}\left(MY\rho Y\right),\\
    r_z &\equiv& \mbox{tr}\left(MZ\rho Z\right),\\
    r_{(x+y)} &\equiv& \mbox{tr}\left(M \left(\frac{X+Y}{\sqrt{2}}\right) \rho  \left(\frac{X+Y}{\sqrt{2}}\right)\right),\\
    r_{(x+z)} &\equiv& \mbox{tr}\left(M \left(\frac{X+Z}{\sqrt{2}}\right) \rho  \left(\frac{X+Z}{\sqrt{2}}\right)\right),\\
    r_{(y+z)} &\equiv& \mbox{tr}\left(M \left(\frac{Y+Z}{\sqrt{2}}\right) \rho  \left(\frac{Y+Z}{\sqrt{2}}\right)\right)\label{eq:relement_stop}.
\end{eqnarray}

Equation~\eqref{eqn:lin-axis} can be simply written as:
\begin{equation}\label{eq:Rcb}
     \left(\sin^2{(\theta/2)}\mathbf{R} - 2\lambda^* \mathbf{I}\right) \hat{\bm{n}}^* = -\alpha_\theta \mathbf{b},
\end{equation}
where $\mathbf{b} \equiv ( \mbox{tr}\left(M\left[\rho, X\right]\right), \mbox{tr}\left(M\left[\rho, Y\right]\right), \mbox{tr}\left(M\left[\rho, Z\right]\right))^T$, and

\begin{flalign}\label{eq:R-def}
    \mathbf{R} &\equiv 
    2\begin{pmatrix}
    r_x  & r_{(x+y)} - \frac{r_x + r_y}{2} & r_{(x+z)} - \frac{r_x + r_z}{2}\\
    r_{(x+y)} - \frac{r_x + r_y}{2} & r_y & r_{(y+z)} - \frac{r_y + r_z}{2}\\
    r_{(x+z)} - \frac{r_x + r_z}{2} & r_{(y+z)} - \frac{r_y + r_z}{2} & r_z
    \end{pmatrix}
\end{flalign}

We can clearly see that $\mathbf{R}$ is a symmetric matrix and therefore its eigenvalues are real numbers. 
Moreover, provided that $\sin(\frac{\theta}{2})\cos(\frac{\theta}{2}) \neq 0$ there exists $\lambda \in \mathbb{R}$ such that~\eqref{eq:Rcb} has a unique solution.
In fact, the value of $\lambda^*$ is the solution to the following identity:
\begin{equation*}%
    1 = \left(\hat{\bm{n}}^*\right)^T \hat{\bm{n}}^* = \left(\alpha_\theta\right)^2 \mathbf{b}^T \left(\left(\sin^2(\theta/2) \mathbf{R} - 2\lambda \mathbf{I}\right)^{-1} \right)^2 \mathbf{b}.
\end{equation*}

Equation~\eqref{eq:Rcb} gives a method to determine an optimal axis of rotation against a fixed $\theta$ by estimating the matrix $\bm{R}$ and vector $\bm{b}$.
We notice that we can further simplify the equation when $\theta = \pi$ that corresponds to replacing the single-qubit gate with a $\pi$-rotation gate around a particular axis.
Although this type of gates is less general than an arbitrary rotational gate, we will see in later sections that not only its expressibility is higher, but also its computational cost to optimize the parameter is better than other ansatz. 
 
\subsection{Simplified Free-Axis Selection}
We have derived a method to select a rotational axis minimizing the objective function by fixing $\theta$.
We refer to it as $\theta$-Fraxis.
The method requires us to run the PQC nine times to estimate the elements of $\mathbf{R}$ and $\mathbf{b}$, which is more than previous methods.
However, we can reduce the number of PQC runs by fixing $\theta = \pi$ as we do not need $\mathbf{b}$ in this case.
Moreover, the unit vector $\hat{\bm{n}}^*$ becomes an eigenvector of $\mathbf{R}$.
The axis that minimizes the objective function is exactly the eigenvector whose eigenvalue is the smallest.
We summarize the $\pi$-Fraxis in Algorithm~\ref{alg:sar}. The stopping criterion can be either reaching the maximum of the number of repetitions or the lack of progress in reaching lower energy values. In the hereafter, for ease of notation we simply call it Fraxis and omit the prefix $\pi$-.  
\begin{figure}[!t]
 \begin{algorithm}[H]
	\caption{Free-axis selection (Fraxis)}\label{alg:sar}
	\begin{algorithmic}
	\Input~A Hermitian matrix $M$ of the objective function, a PQC with single-qubit gates $U_1\ldots U_D$, and a stopping criterion
	\EndInput
	\State \textbf{Initialize:}~Choose random axes, i.e., $\hat{\bm{n}}_d \in \mathbb{R}^3$ such that $|\hat{\bm{n}}_d| = 1$ for $d=1, 2, \ldots, D$. 
	\Repeat
	    \For{d = 1, 2, \ldots D}
	    \State Fix all axes $\hat{\bm{n}}_j$ for $j\neq d$
	    \State Compute elements of  $\mathbf{R}_d$ as in~\eqref{eq:relement_start}--\eqref{eq:relement_stop}
	    \State Compute the eigenvectors $\bm{r}_d^{(1)}, \bm{r}_d^{(2)}, \bm{r}_d^{(3)}$ of $\mathbf{R}_d$
	    \State $\hat{\bm{n}}_d \leftarrow \arg\min_{\bm{r} \in \{\bm{r}_d^{(1)}, \bm{r}_d^{(2)},\bm{r}_d^{(3)}\}} \left<M\right>_{\bm{r}, \pi}$
	    \EndFor
	\Until{satisfying stopping criterion}
	\Output ~Axes of rotation $\left(\hat{\bm{n}}_1, \ldots, \hat{\bm{n}}_D\right)$
	\EndOutput
	\end{algorithmic}
 \end{algorithm}
\end{figure}

We note an important and nice property of the Fraxis: its computational resources turn out to be less than those of Rotoselect~\cite{Ostaszewski2021Quantum} despite the fact that it selects axis of rotation in continuous manner.
While Rotoselect requires evaluating seven energy estimations with the PQC, at each for-loop of Algorithm~\ref{alg:sar} Fraxis needs exactly six energy estimations to compute all elements of $\bm{R}_d$.
The landscape of $\left<M\right>_{\hat{\bm{n}}_d,\pi}$ as in~(\eqref{eqn:obj-H-theta}) is completely determined by the eigenvalues and eigenvectors of $\bm{R}_d$.
We record this property in the following lemma. 

\begin{lemma}\label{lemma:energy-landscape}
Let $\lambda_d^{(i)}$ and $\bm{r}_d^{(i)}$ be, respectively, the $i$-th eigenvalue and eigenvector of $\bm{R}_d$ at each for-loop of Algorithm~\ref{alg:sar} for $i=1,2,3$ so that $\lambda_d^{(1)} \le \lambda_d^{(2)} \le \lambda_d^{(3)}$. Then, for all unit vector $\bm{r} \in \mathbb{R}^3$ it holds that $\lambda_d^{(1)}/2 \le \left<M\right>_{\bm{r}, \pi} \le \lambda_d^{(3)}/2$, and therefore $\hat{\bm{n}}_d = \bm{r}_d^{(1)}$.  
\end{lemma}
\begin{proof}
It follows directly from transforming~(\eqref{eqn:obj-H-theta}) to obtain $\left<M\right>_{\bm{r},\pi} = (1/2)~\bm{r}^T \cdot \bm{R}_d \cdot \bm{r}$, and therefore the minimum (maximum) value is achieved when $\bm{r}$ is the eigenvector whose eigenvalue is minimum (maximum). 
\end{proof}

\begin{figure}[tb]
  \centering
  \includegraphics[scale=0.3]{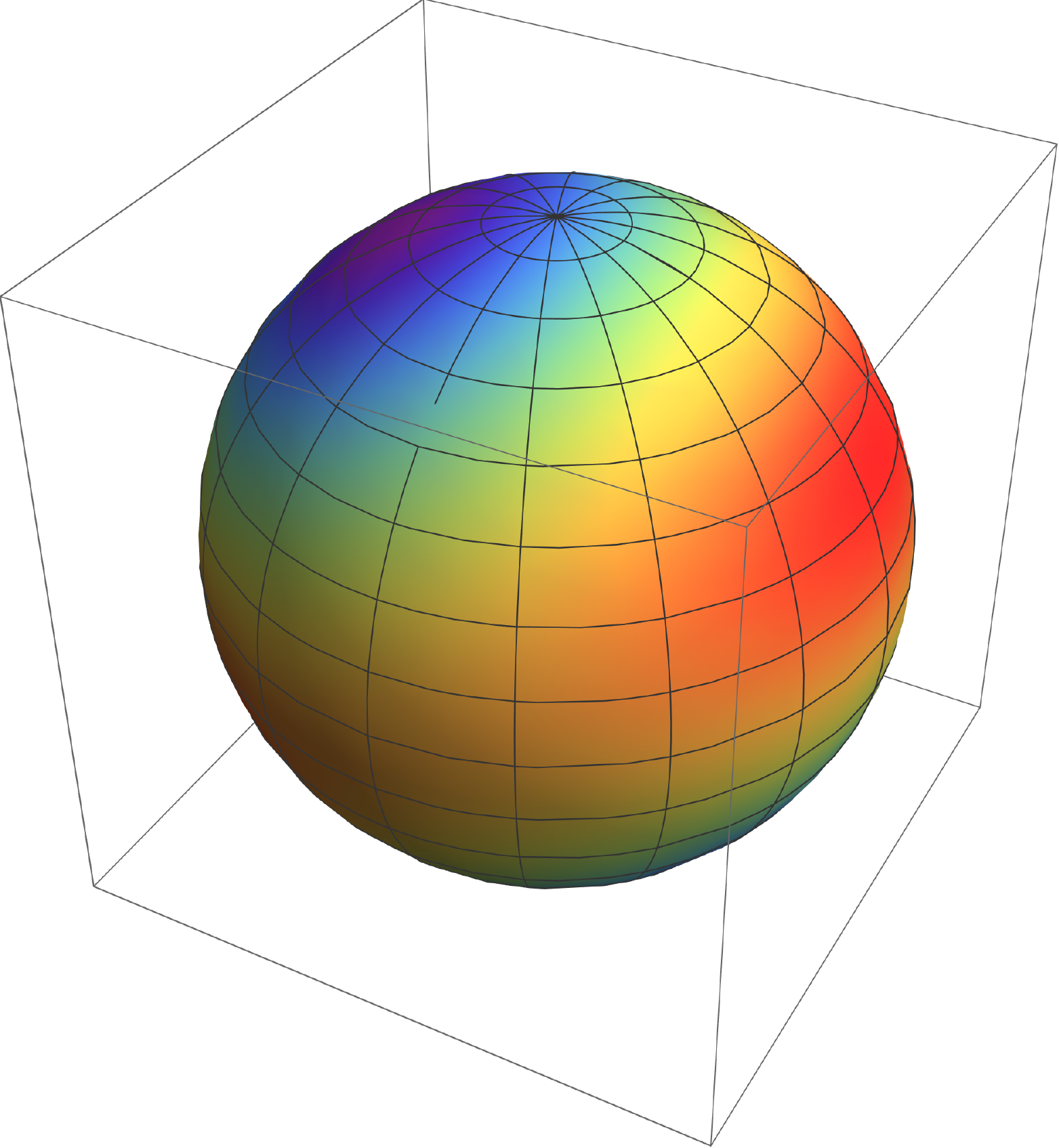}
  \includegraphics[scale=0.3]{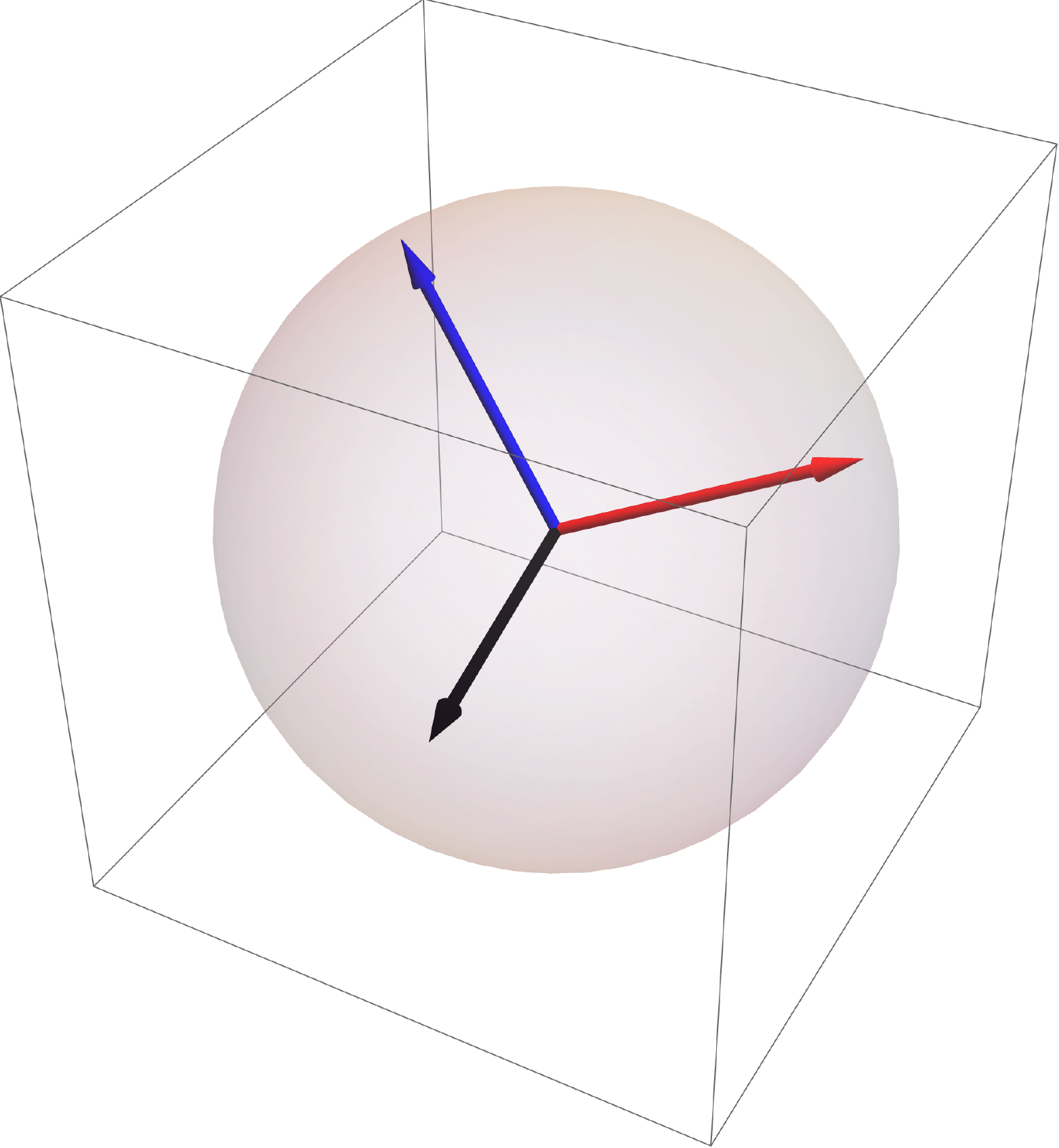}
  \caption{
   (left) An illustration of the energy landscape of Fraxis determined by the matrix $\bm{R}_d$.
   Red, green and blue areas correspond to high, medium, low energy values. 
   (right) Three eigenvectors corresponding to minimum (blue), maximum (red) and saddle (black) points.}
  \label{fgr:EnergyLandscape}
\end{figure}
Figure \ref{fgr:EnergyLandscape} shows the relation between landscape of $\left<M\right>_{\hat{\bm{n}}_d,\pi}$ and three eigenvectors $\bm{r}_d^{(1)}, \bm{r}_d^{(2)}, \bm{r}_d^{(3)}$ 
drawn for simple model example. Eigenvectors denoted by blue and red arrows corresponds to the rotation axes for minimum and maximum energy values, respectively. 

We remark that the cost of energy estimations can be significantly reduced for special Hamiltonians (e.g., $k$-local Hamiltonians~\cite{Kitaev:2002:CQC:863284,Kempe:2006:CLH:1122722.1122821,Oliveira:2008:CQS:2016985.2016987}) if we use locally-biased classical shadows \cite{Hadfield2020arXiv}.
Also, fixing the rotation angle of local gates could be advantageous feature for device control, such as simplification of microwave control pulse calibration for superconducting qubits.
Some potential drawbacks of the Fraxis method are its reduced expressibility (such as, lacking ability to represent identity gates), and its overfitting against the $R_y$ ansatz circuits that are popular for molecular Hamiltonians because such ansatz are sufficient for searching quantum states with real-valued probability amplitudes. We will see later those drawbacks are not significant. 

We also remark that conventional stochastic and gradient approaches may still be compelling for the global optimization in certain cases, such as, the paralellization of variational quantum eigensolvers~\cite{TillyetalVQE2021}. Fraxis can easily be combined with those approaches. For instance, we can analytically evaluate the gradient of the cost functions using~\eqref{eq:dfdx} -- \eqref{eq:dfdz}. The obtained gradient is then passed to the conventional gradient-based optimizers. For this reason, unlike Rotosolve that must update the gates sequentially, Fraxis can be used in the parallelization of variational quantum algorithms, and therefore is potential for avoiding the drawbacks of Rotosolve as mentioned in~\cite{TillyetalVQE2021}. In addition, Fraxis has high affinity with stochastic optimizations, because it gives the exact energy landscape of the single qubit rotation accessible. This can be used to sample the rotation axis at random according to the estimated energy instead of choosing the (local) minimum one.

Two $\pi$-Fraxis gates can realize an arbitrary rotation gate shown in \eqref{eqn:general-rotation} as in the lemma below. This is tight since the identity gate requires two $\pi$-Fraxis gates.

\begin{lemma}
Let an arbitrary single-qubit gate as in~\eqref{eqn:general-rotation} with parameters $(\hat{\bm{n}}, \theta)$. Then, $R_{\hat{\bm{n}}}(\theta) = R_{\hat{\bm{n}}_1}(\pi) R_{\hat{\bm{n}}_2}(\pi)$ provided $\hat{\bm{n}}_1$ and $\hat{\bm{n}}_2$ satisfy $\hat{\bm{n}} = \hat{\bm{n}}_1 \times \hat{\bm{n}}_2$ and $\theta = 2\arccos{(-\hat{\bm{n}}_1\cdot\hat{\bm{n}}_2)}$, where $\times$ and $\cdot$ denote, respectively, the outer and inner products. 
\end{lemma}
\begin{proof}
By~\eqref{eqn:general-rotation}, we can easily see that 
$$
R_{\hat{\bm{n}}_i}(\pi) = -i (n_{i,x} X + n_{i,y} Y + n_{i,z}Z) \equiv -i \hat{\bm{n}}_i\cdot \hat{\sigma},
$$
where $\hat{\sigma} \equiv (X, Y, Z)$ is a vector of Pauli matrices $X, Y$ and $Z$. Utilizing the above identity, we can confirm that 
\begin{eqnarray*}
R_{\hat{\bm{n}}_1}(\pi) R_{\hat{\bm{n}}_2}(\pi) &=&  - (\hat{\bm{n}}_1 \cdot \hat{\bm{n}}_2) I - i  \left( (\hat{\bm{n}}_1 \times \hat{\bm{n}}_2) \cdot \hat{\sigma} \right)\\
&=& R_{\hat{\bm{n}}_1 \times \hat{\bm{n}}_2} \left(2\arccos(-\hat{\bm{n}}_1\cdot \hat{\bm{n}}_2) \right)
\end{eqnarray*}
\end{proof}

We end this section with an important note on the implementability of $\pi$-Fraxis gates on IBM Quantum devices that do not natively support Fraxis gates. The $\pi$-Fraxis gate is decomposed into native single-qubit gates as follows.
\begin{equation}
 R_{\hat{\bm{n}}}(\pi) = R_{z}(\pi + \phi)\cdot \sqrt{X} \cdot R_z(\pi+\theta) \cdot \sqrt{X}\cdot R_z(\pi - \phi),  
\end{equation}~\label{eqn:fraxis-decomposed}
where $\theta = 2\arctan\left(\sqrt{n_x^2+n_y^2}/{n_z}\right)$ and $\phi = \arctan(n_y/n_x)$. Because the $R_z$ gates on IBM Quantum devices do not require \textit{pulse} operation~\cite{McKay2017}, the $\pi$-Fraxis gate can be realized with the actual cost of two $\sqrt{X}$ gates, which is essentially the same as the $R_yR_z$ gate. We can also notice from the decomposition that the $\pi$-Fraxis gate has two degrees of freedom, which is also the same as the $R_yR_z$ gate. Optimizing the $R_yR_z$ gate with Rotosolve~\cite{Nakanishi2020PRR,Ostaszewski2021Quantum} proceeds with sequentially optimizing the parameters of $R_y$ and $R_z$ one at a time, while the $\pi$-Fraxis gate optimizes the two parameters simultaneously. There is a discussion on how to optimize correlated parameters with Rotosolve but, unlike the $\pi$-Fraxis, there is no known closed-form solution~\cite{Nakanishi2020PRR}.

%% file: result_body.tex
In this section, we first apply $\theta$-Fraxis to simple 2-qubit system and compare it to Rotosolve and Rotoselect, and show its advantage in energy convergence, together with the potential of the simplified $\pi$-Fraxis. 
Next, using Qiskit~\cite{Qiskit} we demonstrate the high expressibility of PQCs with the simplified $\pi$-Fraxis (or, Fraxis) in contrast to PQCs with its traditional counterparts.
For studying expressibility, we show the one-qubit PQCs with Fraxis can generate quantum states covering the Bloch sphere more uniformly that its counterparts. 
We then show evidences of better expressibility of multi-qubit PQCs with Fraxis when measured in KL divergences under different number of entangled layers.
Finally, we present numerical experiments on simulators and near-term quantum devices showing Fraxis can further improve the efficacy of variational quantum circuits for the Heisenberg model, molecular, and combinatorial optimization Hamiltonians. 

\subsection{Comparison with Other Structural Optimization}\label{subsec:com_opt}
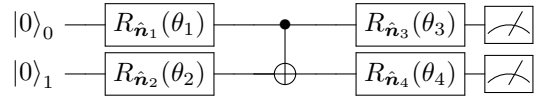
\begin{figure}[tb]
\[
\Qcircuit @C=.75em @R=.25em {
  \lstickx{\ket{0}_0} & \qw & \gate{\Rn[1]} & \qw & \qw & \ctrl{1} & \qw  & \qw& \gate{\Rn[3]}  & \meter \\
  \lstickx{\ket{0}_1} & \qw & \gate{\Rn[2]} & \qw & \qw & \targ \qw & \qw  & \qw& \gate{\Rn[4]}  & \meter \\
}
\]
  \caption{A 2-qubit PQC with $\theta$-Fraxis.}\label{fgr:2qubit_PQC}
\end{figure}

For comparison, we tested Rotosolve, Rotoselect, $\theta$-Fraxis, and Fraxis methods to a 2-qubit PQC system in Fig.~\ref{fgr:2qubit_PQC} for finding the ground state of the Hamiltonian
\begin{equation}
\label{2qubit-model}
    H= 0.1 (XX +YY + ZZ) + 0.01(IZ + ZI). 
\end{equation}
The ground state energy is $E_0 = -0.3$.
The variational parameters to be optimized are four rotation angles $(\theta_1, \theta_2, \theta_3, \theta_4)$ and their corresponding rotational axes, $\left(\hat{\bm{n}}_1, \hat{\bm{n}}_2, \hat{\bm{n}}_3, \hat{\bm{n}}_4\right)$.
We applied the methods to start from the same 50 random initial parameter sets.

\begin{figure*}[tb]
  \centering
  \begin{tabular}{cc}
   \includegraphics[scale=0.40]{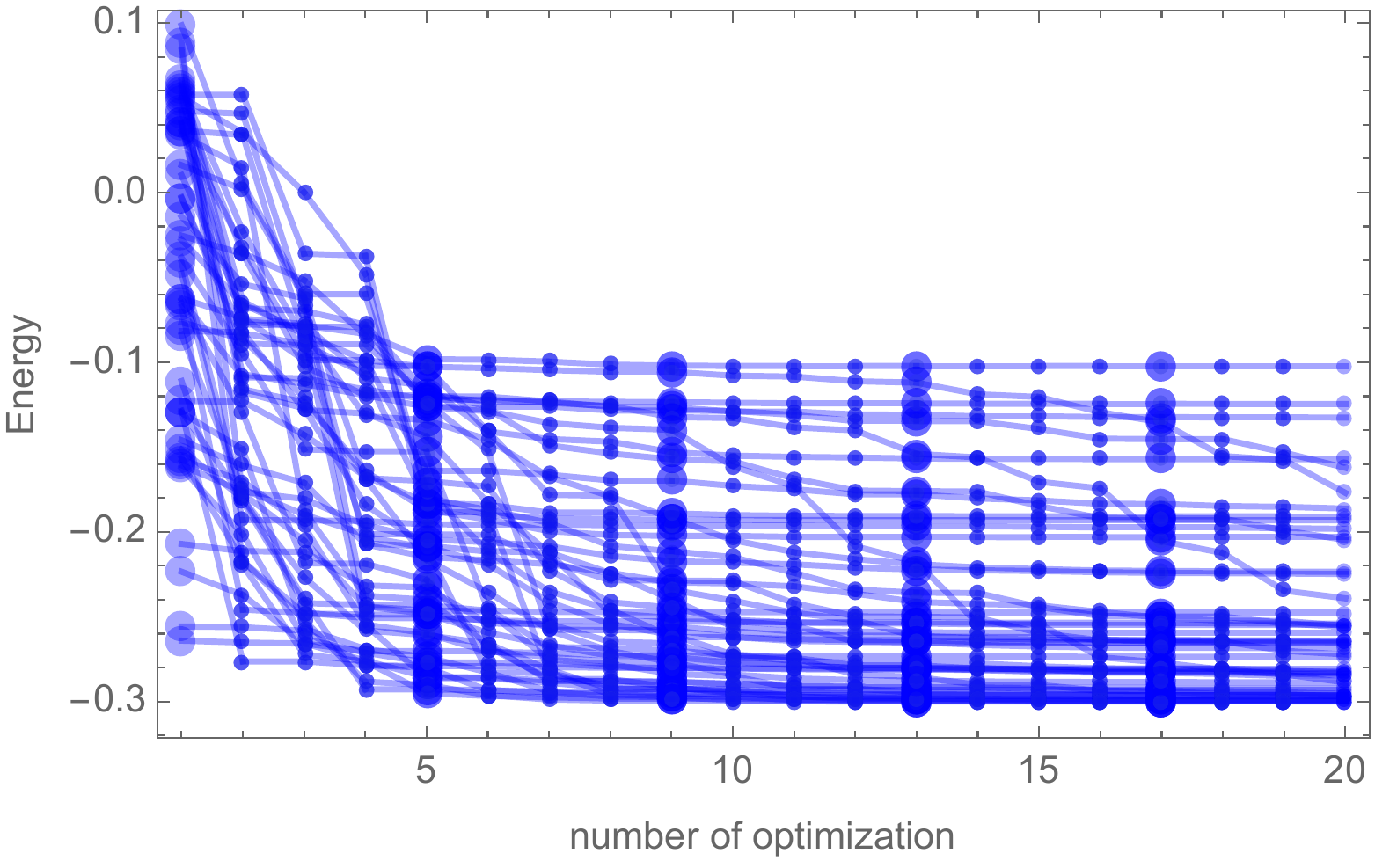} & 
   \includegraphics[scale=0.40]{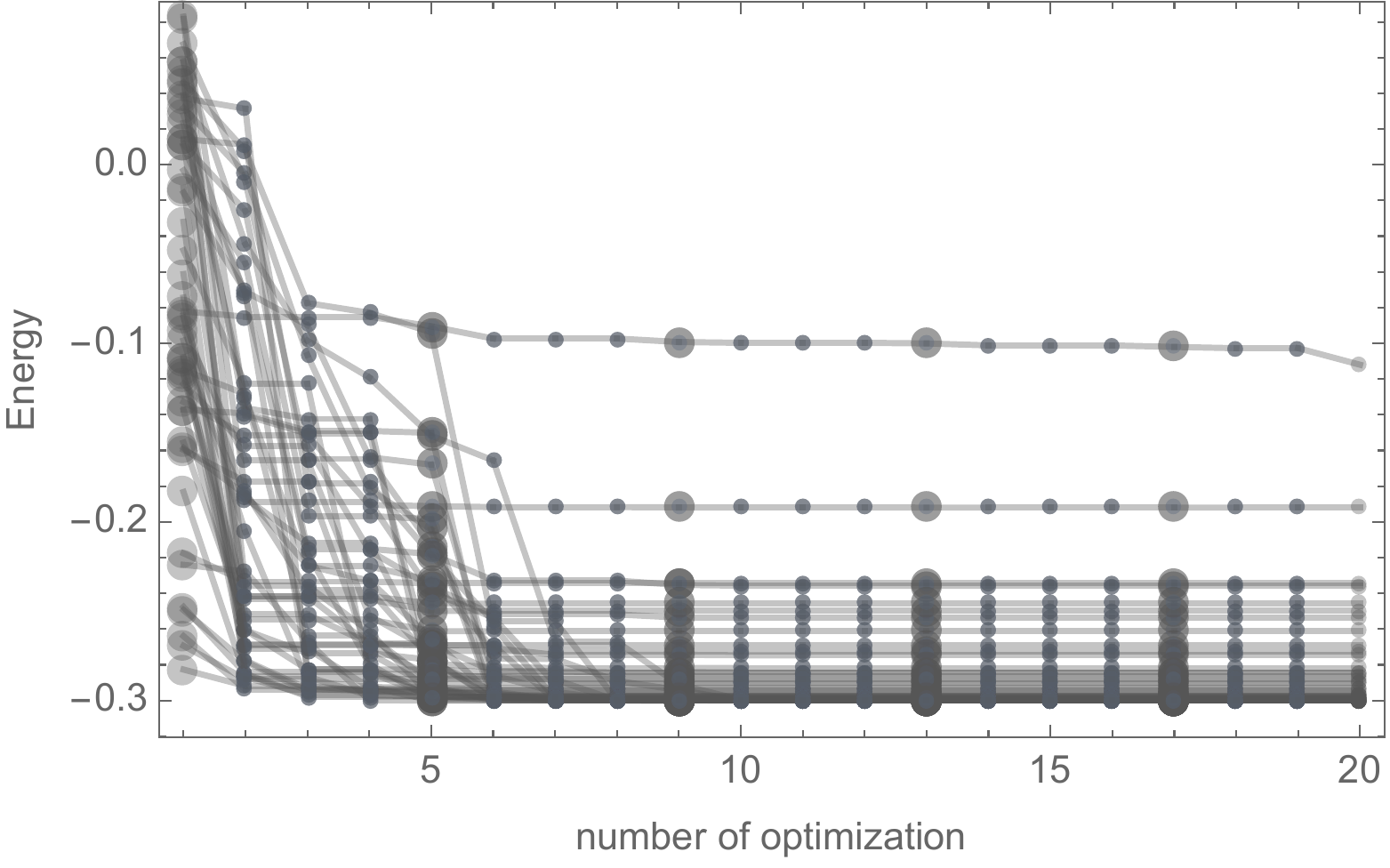}\\
   (a) Rotosolve &
   (b) Rotoselect \\
   \includegraphics[scale=0.40]{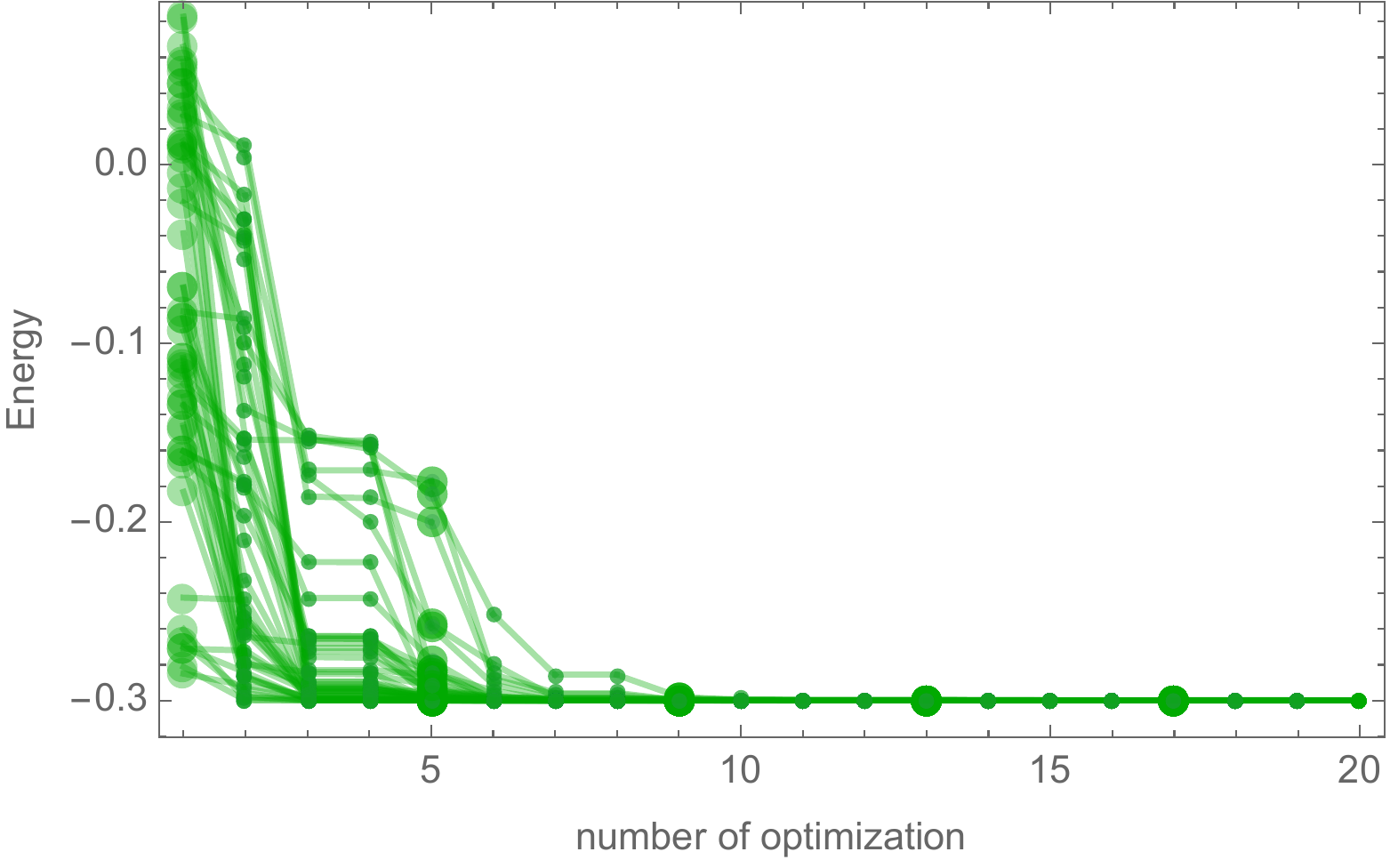} &
   \includegraphics[scale=0.40]{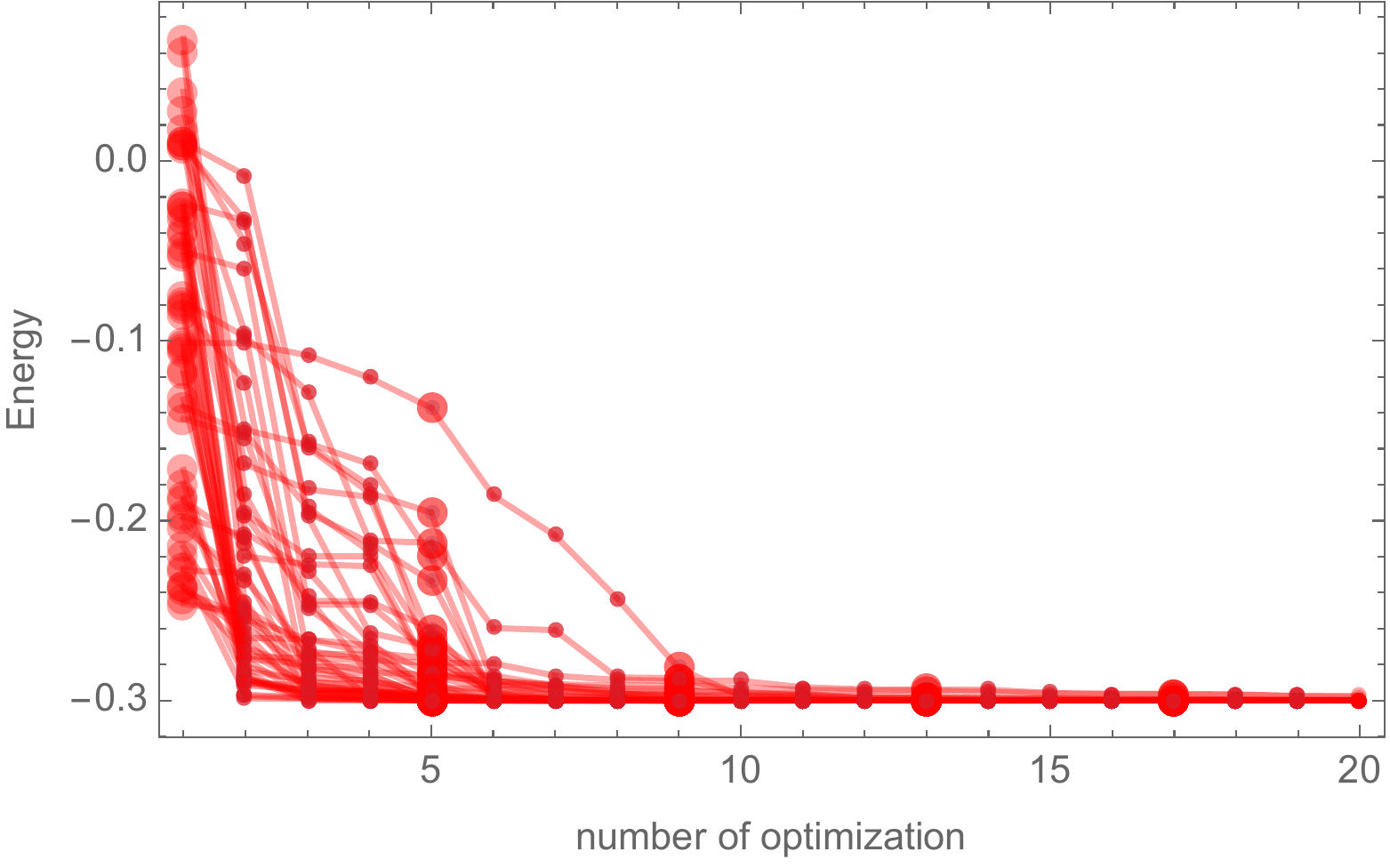} \\
   (c) $\theta$-Fraxis &
   (d) $\pi$-Fraxis (Fraxis)\\
  \end{tabular}
  \caption{Comparison of energy convergence as a function of number of iterations for optimization. 
  (a) Rotosolve (blue), (b) Rotoselect (gray), (c) $\theta$-Fraxis (green), and (d) Fraxis (red). 
  Common initial angular parameter set is used for Rotosolve, Rotoselect and $\theta$-Fraxis, 
  while $(\theta_1, \theta_2, \theta_3, \theta_4)$ are overridden to $(\pi, \pi, \pi, \pi)$ in Fraxis trials.
  }
  \label{fgr:comparison}
\end{figure*}
Figure~\ref{fgr:comparison} shows the optimized energy as a function of iterative number of optimization.
One can see that $\theta$-Fraxis shows better convergence compared to Rotosolve or Rotoselect, while those methods seem to suffer from being trapped in local minima. 
This is because Rotosolve and Rotoselect mainly optimize rotation angles $(\theta_1, \theta_2, \theta_3, \theta_4)$ and leave rotation axes unchanged (Rotosolve) or selected from the limited choice, $R_x$, $R_y$ or $R_z$ (Rotoselect). 
Although Rotoselect result exhibits better convergence nature, i.e., less likely to be trapped in local minima when compared to Rotosolve, some parameter sets could not lead to the exact ground state energy $E_0$.

On the other hand, $\theta$-Fraxis makes it possible to optimize rotation axis continuously with respect to given $\theta$,  which results in fine convergence nature as shown in Fig.~\ref{fgr:comparison}. 
We also applied the simplified Fraxis with $\theta = \pi$ (or, $\pi$-Fraxis) to the same problem.
Note that the initial parameters for axes are the same in Rotosolve, Rotoselect and $\theta$-Fraxis cases, while rotation angles $(\theta_1, \theta_2, \theta_3, \theta_4)$ are set to $(\pi,\pi,\pi,\pi)$.
The result in Fig.~\ref{fgr:comparison} shows that the convergence behaviour of $\pi$-Fraxis is still better than those of Rotosolve/select.
We also remark that $\pi$-Fraxis has comparable performance to $\theta$-Fraxis.
Nevertheless, estimation cost for matrix-elements in $\pi$-Fraxis is reduced from 9 to 6, which suggests $\pi$-Fraxis is more practical than $\theta$-Fraxis.
Recall that we abbreviate $\pi$-Fraxis as Fraxis and will further investigate expressibility of Fraxis ansatz/circuit to ensure the usability of this simplified method in the next subsection.

\subsection{Circuit Expressibility}\label{subsec:circ-exp}
In general, a universal gate for a single qubit is represented as \begin{equation}\label{eqn:universal-unitary}
    U(\psi, \phi, \lambda)=
    \left[
    \begin{array}{cc}
    \cos{\frac{\psi}{2}} & 
    -e^{i\lambda}\sin{\frac{\psi}{2}} \\
    e^{i\phi}\sin{\frac{\psi}{2}} & e^{i(\phi+\lambda)}\cos{\frac{\psi}{2}} \\ 
    \end{array}
    \right].
\end{equation}

In the case of Fraxis, the rotation angle is fixed at $\theta=\pi$ in~\eqref{eqn:general-rotation}, which is equivalent the constraint condition $\phi+\lambda=\pi$ in~\eqref{eqn:universal-unitary}.
Hence, a single gate controlled by Fraxis is not universal any more. 
However, it should be noted that any input state in a single qubit system can be transformed to an arbitrary state.
To intuitively understand this, let us suppose one qubit state in $\ket{0}$.
This $\ket{0}$ state can be transformed to any state on the Bloch sphere by $\pi$ rotation around the axis that passes through the dividing point of $\ket{0}$ and the target state. 
Therefore, the Fraxis ansatz seems to retain high expressibility. Moreover, it can also be shown that a universal 1-qubit gate in~\eqref{eqn:universal-unitary} can be realized by at most 2 Fraxis gates, and therefore a hard instance of general quantum circuit can be turned into a hard instance of Fraxis ansatz whose size is in the same order of the general quantum circuit. 

To quantitatively evaluate the expressibility of the Fraxis circuit in comparison with Rotosolve and Rotoselect, we evaluate Kullback-Leibler (KL) divergence as
\setlength{\arraycolsep}{0.0em}
\begin{eqnarray}{lCl}
    \mathcal{E(C)} 
    &{}={}& D_{\mathrm{KL}} \left( P(C,F)||P_{\mathrm{Haar}}(F) \right) \nonumber \\ 
    &{}={}&\int_0^1 P(C,F)  \log{\frac{P(C,F)}{P_\mathrm{Haar}(F)}} dF,\label{KL_Divergence}
\end{eqnarray}
\setlength{\arraycolsep}{5pt}
where $F$ is the fidelity between two parametrized random states $\ket{\psi(\theta)}, \ket{\psi(\theta')}$ which is for two pure states defined as $F = |\braket{\psi|\psi'}|^2$.
Here, $P(C, F)$ is a \textit{probability distribution function} (pdf) of fidelity $F$ of a circuit ansatz $C$. 
$P_\mathrm{Haar}(F)$ is a probability distribution sampled uniformly according to the Haar measure, in which $\ket{\psi}$ is uniformly distributed in Hilbert space.
According to~\cite{PhysRevA.71.032313} the Haar-measure distribution is derived as, $P_\mathrm{Haar}(F)= (N-1)(1-F)^{N-2}$, where $N$ is the dimension of Hilbert space. 
Here, a product state of ${\otimes}^n\ket{0}$ is employed as the initial state. 

\begin{figure*}[tb]
 \centering
 \begin{tabular}{ccc}
  \Qcircuit @C=.75em @R=.25em {
  \lstickx{\ket{0}} & \qw & \gate{R_y} & \qw & \meter  & \qw & \qw \\
  } &~&
  \Qcircuit @C=.75em @R=.25em {
  \lstickx{\ket{0}} & \qw & \gate{R_x/R_y} & \qw  & \meter   & \qw     & \qw \\
  } \\
  \includegraphics[scale=0.2]{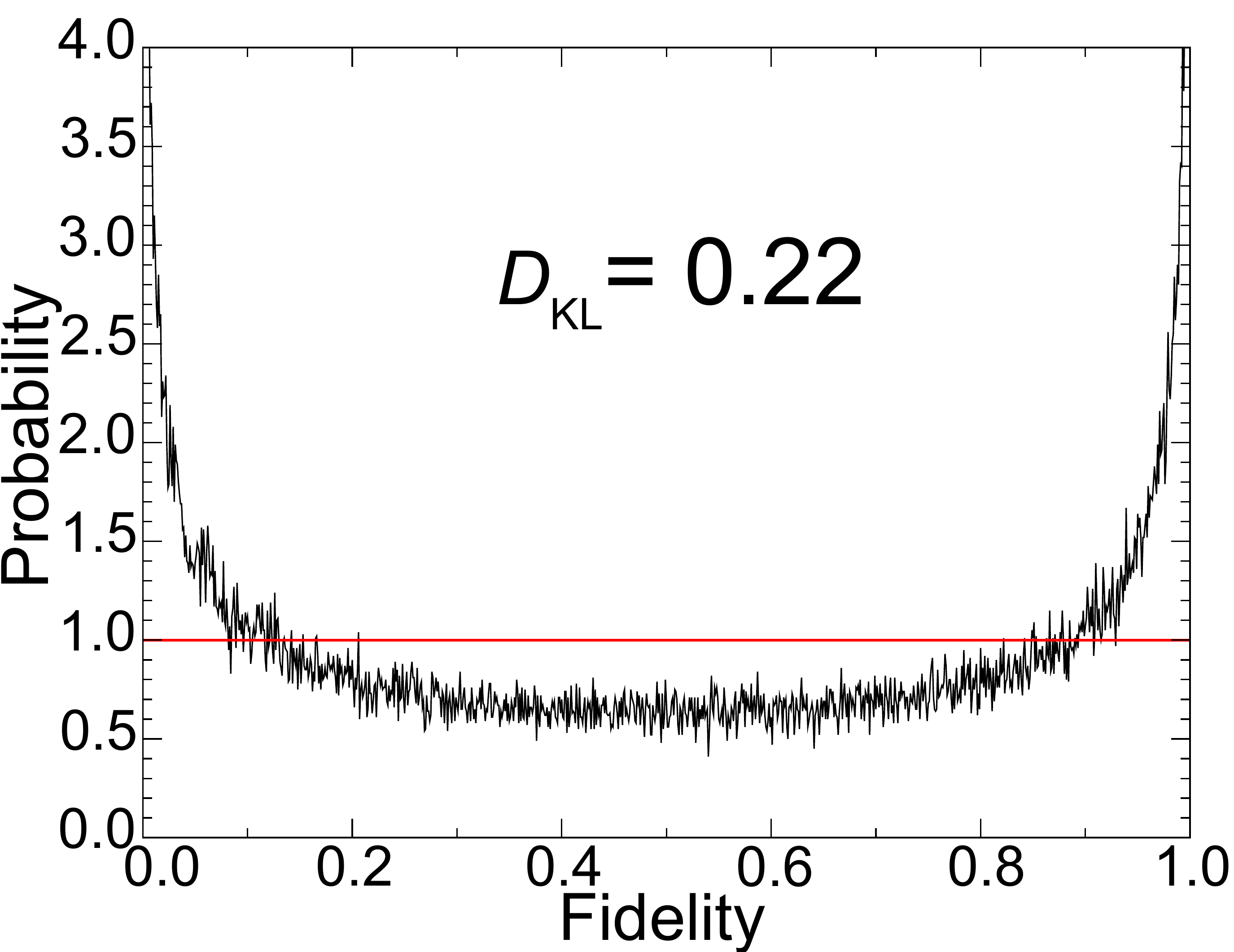} &~&
  \includegraphics[scale=0.2]{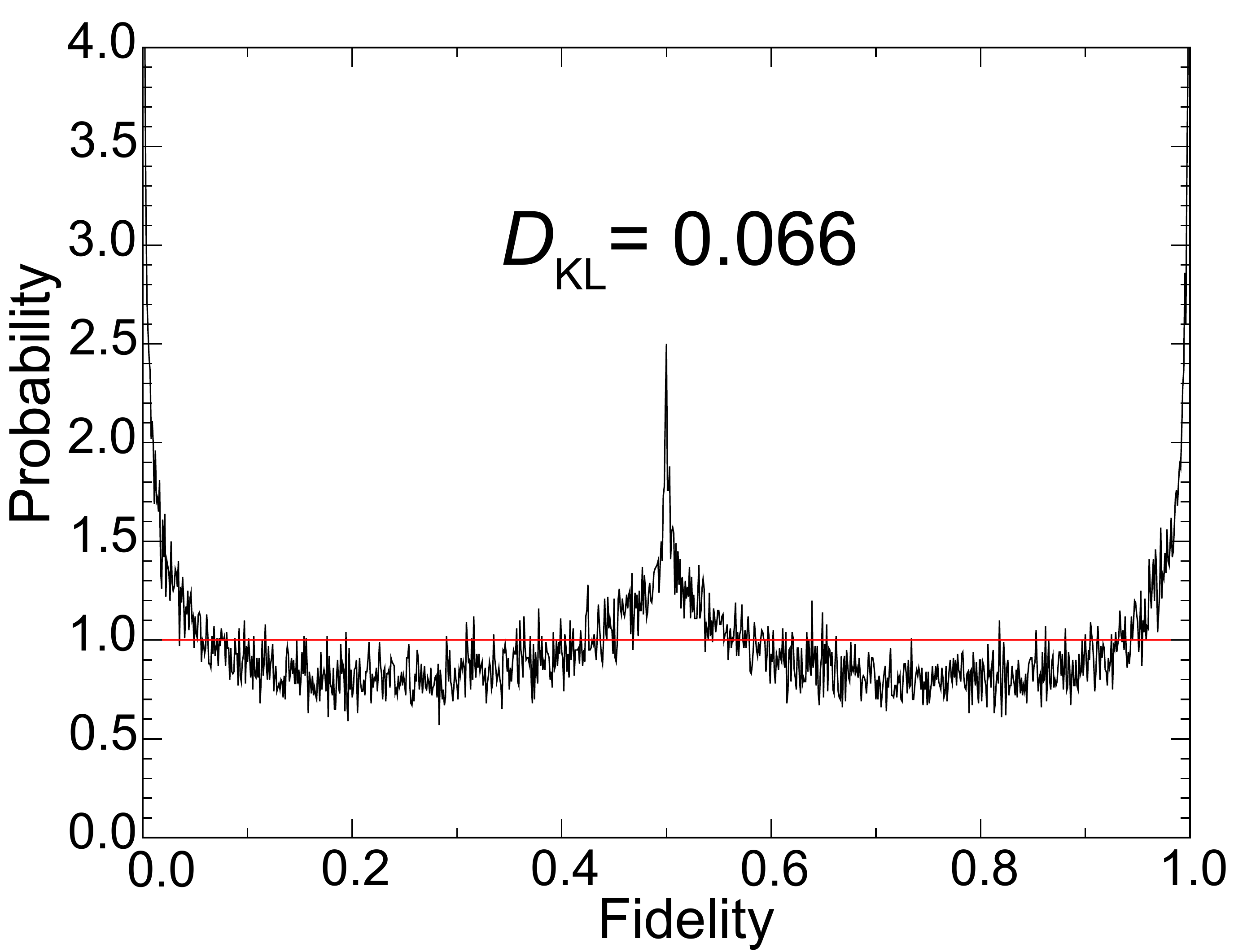}\\
  (a) Rotosolve  &~&
  (b) Rotoselect  
 \end{tabular}
\caption{Fidelity distribution obtained by single-qubit circuits with (a) Rotosolve($R_y$), and (b) Rotoselect($R_x/R_y$). Horizontal lines in red represent Haar-uniform distribution.
The evaluated KL divergence values are displayed over the respective plots}\label{fgr:rotosolve-select}
\end{figure*}

\begin{figure*} [tb]
\centering
\begin{tabular}{ccc}
\multicolumn{3}{c}{\Qcircuit @C=.75em @R=.25em {
  \lstickx{\ket{0}} & \qw & \gate{\UU}  & \qw & \meter   & \qw     & \qw \\
}}\\
\includegraphics[scale=0.2]{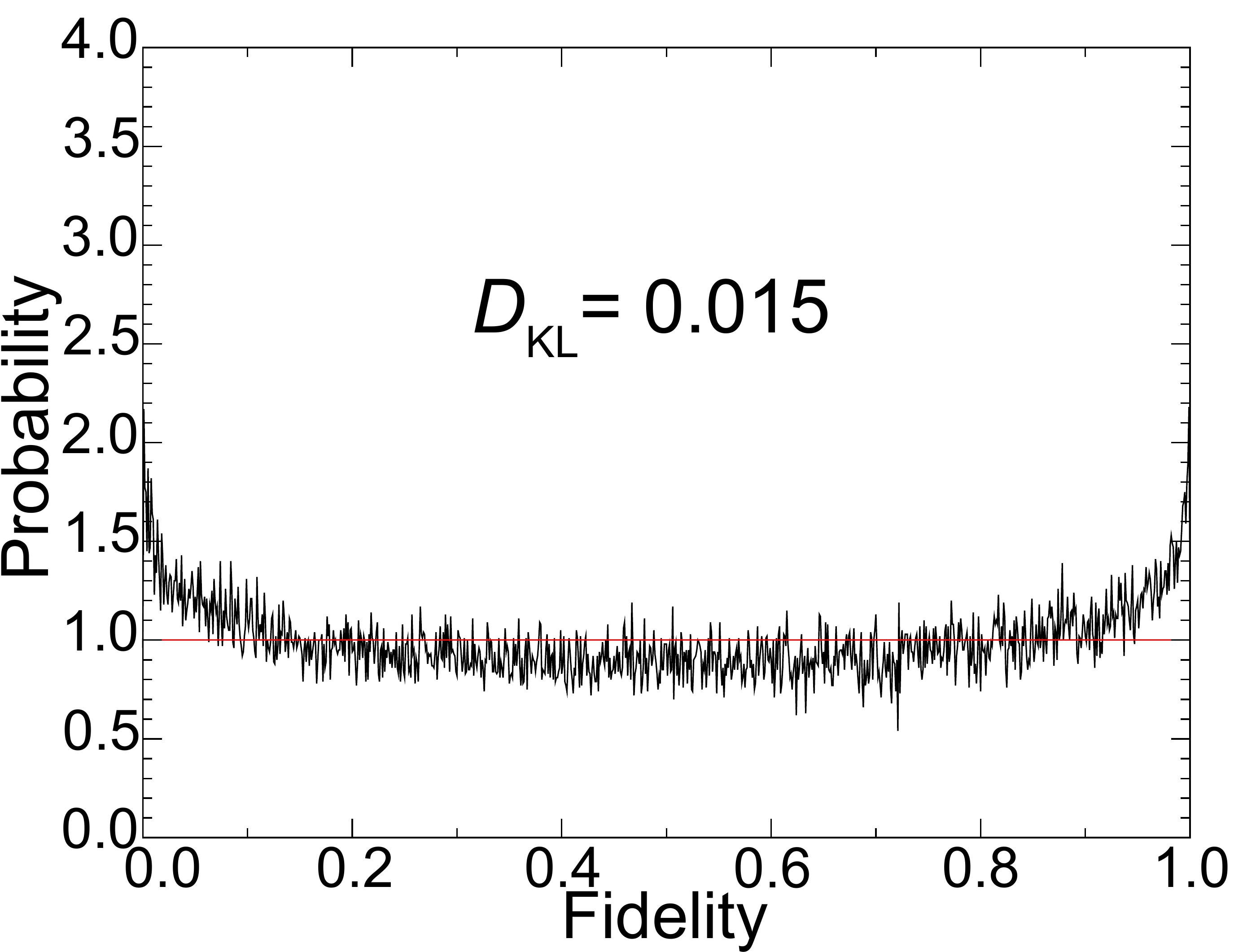} &~&
\includegraphics[scale=0.2]{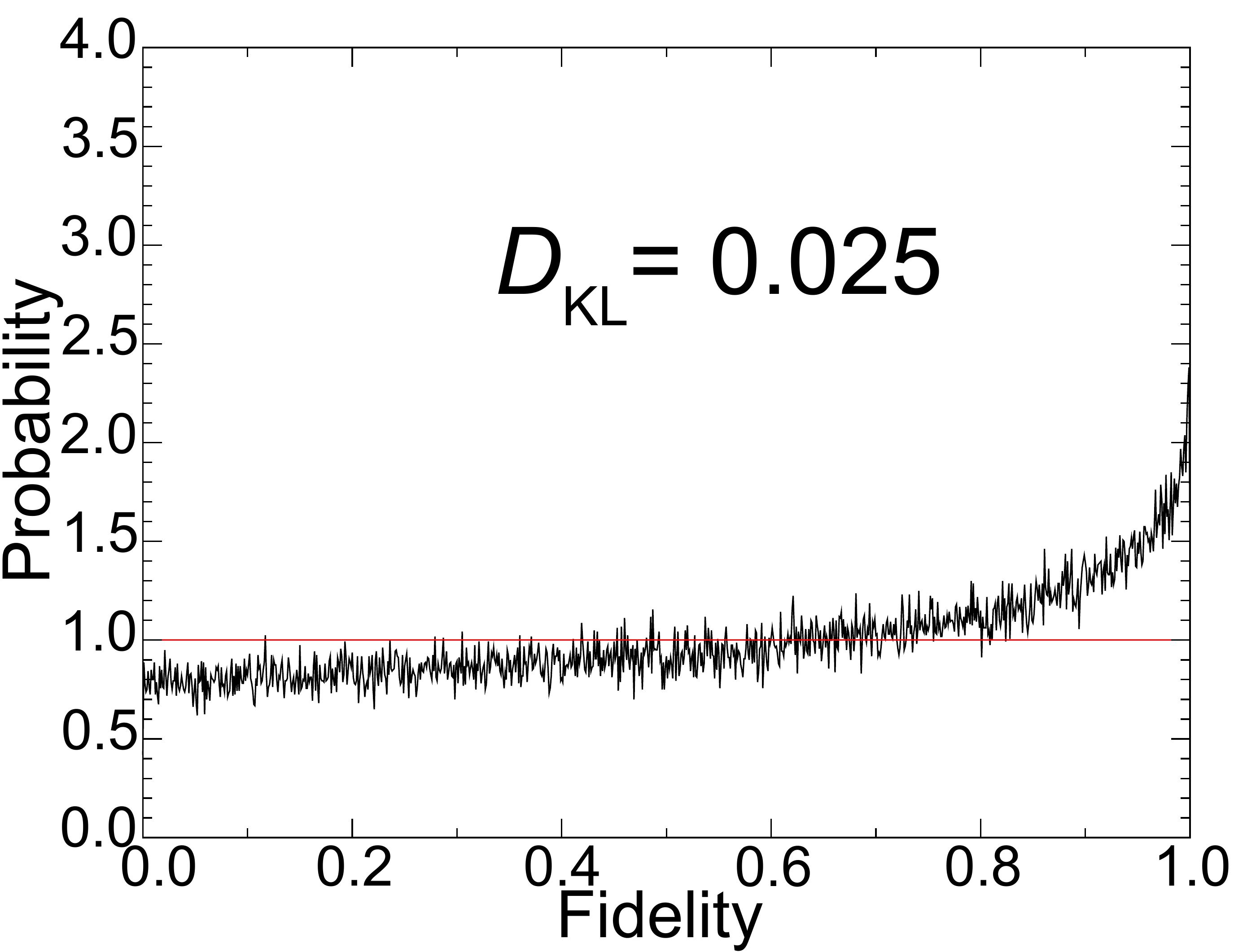}\\
\textbf{(a)} parameter-random Fraxis  &~& \textbf{(b)} state-random Fraxis
\end{tabular}
 \caption{Fidelity distribution obtained by single-qubit circuits with (a) parameter-random Fraxis, and (b) state-random Fraxis. Horizontal lines in red represent Haar-uniform distribution.
 The evaluated KL divergence values are displayed over the respective plots.}\label{fgr:sar-random}
\end{figure*}

First, we confirmed the expressibility of single-qubit circuits as shown in Figs.~\ref{fgr:rotosolve-select}~and~\ref{fgr:sar-random}, where $\ket{0}$ is employed as the initial state. 
Since the rotation axis is fixed in Rotosolve ansatz, the expressibility is equivalent to that of a single $R_y(\theta)$ gate. 
Then, the rotation angles $\theta$ were randomly generated in the range of $(-\pi, \pi]$. 
Due to symmetry, the fidelity is analytically given as $F=f(\theta)=|\braket{0|\psi(\theta)}|^2=\cos^2(\theta/2) $ where $\theta$ is uniformly distributed.
Then, the pdf of the fidelity $P(F)$ can be defined as $P(F) \propto |f^{-1}(F) / dF|$.
Because $f^{-1}(F) = -2 \arccos(F)$, we obtain 
\begin{equation}
P(F) = \frac{1}{\pi\sqrt{F(1-F)}}.\label{PF-func}
\end{equation}
Substituting~\eqref{PF-func} into~\eqref{KL_Divergence} gives theoretical value of $D_{KL}=\log (4/\pi)=0.24$ for Rotosolve with $R_y$-ansatz.
As mentioned in the previous study~\cite{Sim2019AQT}, we confirmed the fact that the KL divergence is largely sensitive to the bin width of the pdf histogram, and thus the proper width is not trivial.
Here, we evaluated the histogram with bin width of 0.001 and 100,000 fidelity sampling and obtained numerically the $D_{KL}=0.22$ for the Rotosolve (as in Fig~\ref{fgr:rotosolve-select}), in close agreement with the analytical value. Notice from the figure that the high value of the KL divergence is due to discrepancies of probability distributions around the edges, i.e., $F=0, 1$ and a wide range centered at $F=0.5$. For fair comparison, all expressibility evaluations are based on bin width of 0.001 and 100,000 sampling throughout this paper.

In the Rotoselect circuit, the rotation gates are usually selected from $R_x(\theta)$, $R_y(\theta)$ and $R_z(\theta)$. 
Since the $R_z(\theta)$ rotation does not change the initial state $\ket{0}$, we here randomly selected either $R_x(\theta)$ or $R_y(\theta)$ with equal probabilities as well as the rotation angle that uniformly distributed in the range of  $(-\pi, \pi]$.
In other words, the resulting pdf of the fidelity is a mixture of those of $|\bra{0}R_y(\theta') R_y(\theta)\ket{0}|^2$ and $|\bra{0}R_x(\theta') R_y(\theta)\ket{0}|^2$.  
While the former term contributes in the same way as Rotosolve, the cross evaluation between $R_x$ and $R_y$ contributes in the peak at $F=0.5$ that in turn lifts the pdf closer to the Haar-random. This gives a better expressibility of Rotoselect ansatz.

In the Fraxis circuits, while the rotation angles are fixed $\theta=\pi$, the rotation axis $\hat{\bm{n}}=(n_x, n_y, n_z)^T$ are generated randomly. 
Since the Fraxis ansatz optimizes the target gate in a deterministic way without systematic search in parameter space, it is not straightforward to compare the expressibility with conventional parametrized circuits.
Indeed, the description of randomness of Fraxis ansatz seems not to be unique and we have to arbitrarily choose to describe randomness in either parameter space (parameter random) or Hilbert space (state random).

In parameter random, the rotation axis is represented as $\hat{\bm{n}}=(\sin\theta \cos\phi, \sin\theta \sin\phi, \cos\theta)^T$ and $\theta$ and $\phi$ are randomly sampled with uniform probability in ranges of $[0, \pi)$ and  $[-\pi, \pi)$, respectively. In state random, each element of rotation axis (i.e., $n_x$, $n_y$, $n_z$) was randomly generated according to the normal distribution, and then normalized by the sum of their squares.

For qualitative understanding of the difference between parameter and state randoms, let us consider the surface area on Bloch sphere.
Here, we can consider only the rotation axis pointing to the north sphere of the Bloch sphere without loss of generality because of the symmetry of $\pi$ rotation.
A set of rotation axis pointing south from of latitude 45 degree transforms $\ket{0}$ to a state in the south sphere on Bloch sphere, while the rest of the axis group transforms $\ket{0}$ to a state in the north sphere.

Since the surface area south from latitude 45 degree is larger than the rest, the uniform distribution of rotation axis on Bloch sphere results in larger population of transformed state on south sphere.
This bias will cause asymmetry of the pdf of fidelity as shown in Fig.~\ref{fgr:sar-random}
In the parameter random, the rotation axis $\hat{\bm{n}}$ is likely to be more populated around north/south pole on the Bloch sphere.
Accordingly, the rotated states are also populated around $\ket{0}$, which increases the probability around $F=0$ and $1$ as shown in Fig~\ref{fgr:sar-random}.
Although the KL divergences vary depending on how the random states are generated in Fraxis, it is notable that both pdfs are almost uniform close to Haar measure even with the fixed rotation angle $\theta=\pi$ for 1-qubit systems. 

\begin{figure*} [tb]
\centering
\begin{tabular}{cccccccc}
\Qcircuit @C=.75em @R=.25em {
  \lstickx{\ket{0}_0} & \qw & \gate{\UU[l+0]} & \ctrl{1}          & \qw              & \qw& \gate{\UU[L+0]}  & \meter   & \qw     & \qw \\
  \lstickx{\ket{0}_1} & \qw & \gate{\UU[l+1]} & \control \qw & \ctrl{1}          & \qw& \gate{\UU[L+1]}  & \meter    & \qw      & \qw \\
  \lstickx{\ket{0}_2} & \qw & \gate{\UU[l+2]} & \ctrl{1}         & \control \qw  & \qw& \gate{\UU[L+2]}  & \meter    & \qw     & \qw \\
  \lstickx{\ket{0}_3} & \qw & \gate{\UU[l+3]} & \control \qw & \ctrl{1}          & \qw& \gate{\UU[L+3]} & \meter   & \qw      & \qw \\
  \lstickx{\ket{0}_4} & \qw & \gate{\UU[l+4]} & \qw              & \control \qw  & \qw& \gate{\UU[L+4]}  & \meter   & \qw      & \qw \\
  & & & & &    \arrep{llll}
  \gategroup{1}{3}{5}{5}{.7em}{--}
}
&~&~&~&~&~&~&
\Qcircuit @C=.75em @R=.25em {
  \lstickx{\ket{0}_0} & \gate{X}& \qw & \gate{\UU[l+0]} & \ctrl{1}          & \qw              & \ctrl{3}          & \qw  & \gate{\UU[L+0]}  & \meter   & \qw     & \qw \\
  \lstickx{\ket{0}_1} & \qw       & \qw & \gate{\UU[l+1]} & \control \qw  &\ctrl{1}          & \qw               & \qw  & \gate{\UU[L+1]}  & \meter    & \qw      & \qw \\
  \lstickx{\ket{0}_2} & \qw       & \qw & \gate{\UU[l+2]} & \qw               & \control \qw  & \qw              & \qw  & \gate{\UU[L+2]}  & \meter    & \qw     & \qw \\
  \lstickx{\ket{0}_3} & \gate{X}& \qw & \gate{\UU[l+3]} & \ctrl{1}          & \qw              & \control \qw  & \qw  & \gate{\UU[L+3]} & \meter   & \qw      & \qw \\
  \lstickx{\ket{0}_4} & \qw       & \qw & \gate{\UU[l+4]} & \control \qw  &\ctrl{1}          & \qw               & \qw  & \gate{\UU[L+4]}  & \meter   & \qw      & \qw \\
  \lstickx{\ket{0}_5} & \qw       & \qw & \gate{\UU[l+5]} & \qw               & \control \qw  & \qw              & \qw  & \gate{\UU[L+5]}  & \meter   & \qw      & \qw \\
  & & & & & &  &  \arrep{lllll}
  \gategroup{1}{4}{6}{7}{.7em}{--}
}\\
&~&~&~&~&~&~& \\
\textbf{(a)} 5-qubit circuit A  &~&~&~&~&~&~& \textbf{(b)} 6-qubit circuit B
\end{tabular}
\caption{PQCs with Fraxis-ansatz}\label{fgr:circuits}
\end{figure*}

\begin{figure*}[tb]
  \centering
  \begin{tabular}{cc}
  \includegraphics[scale=0.25]{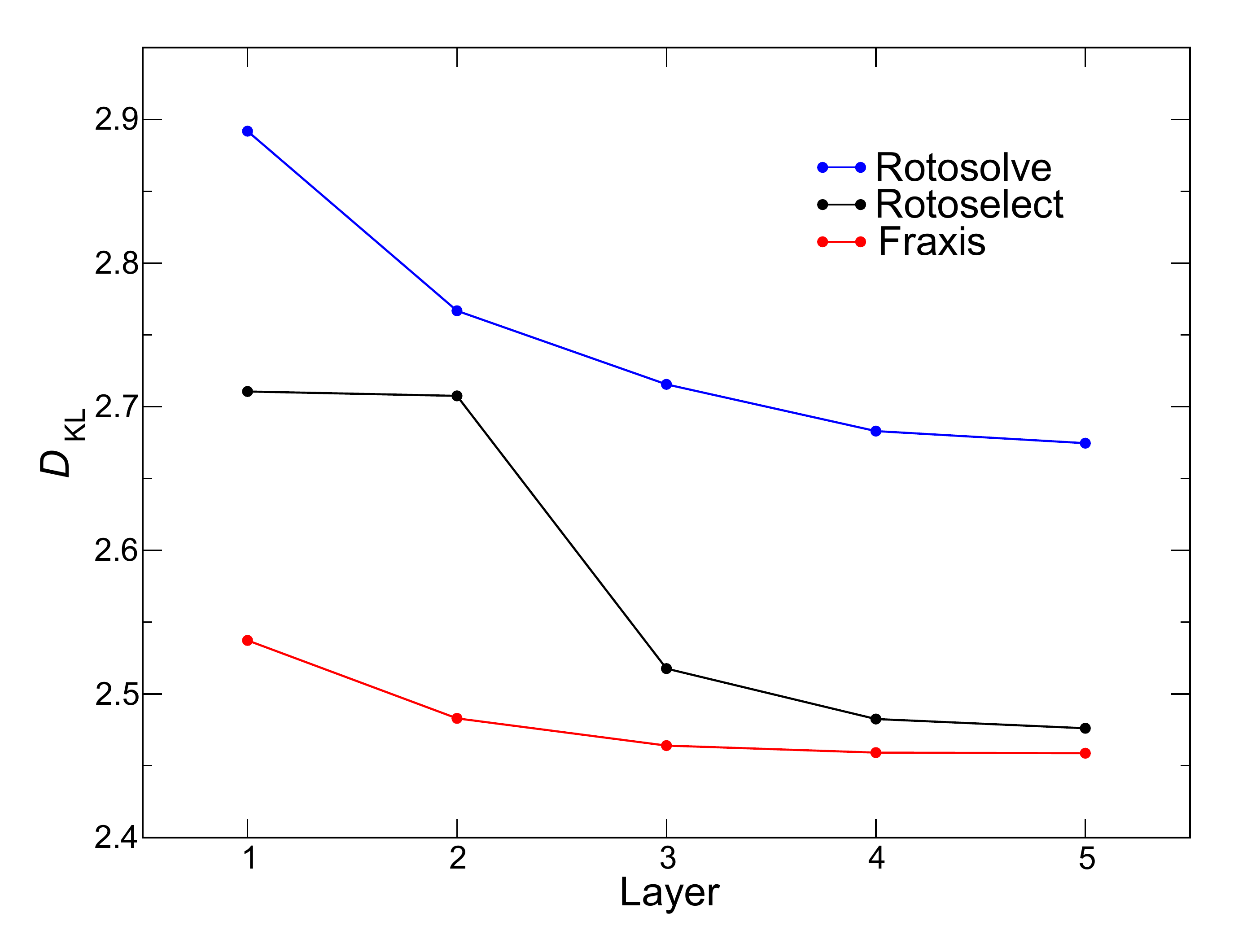} & 
  \includegraphics[scale=0.25]{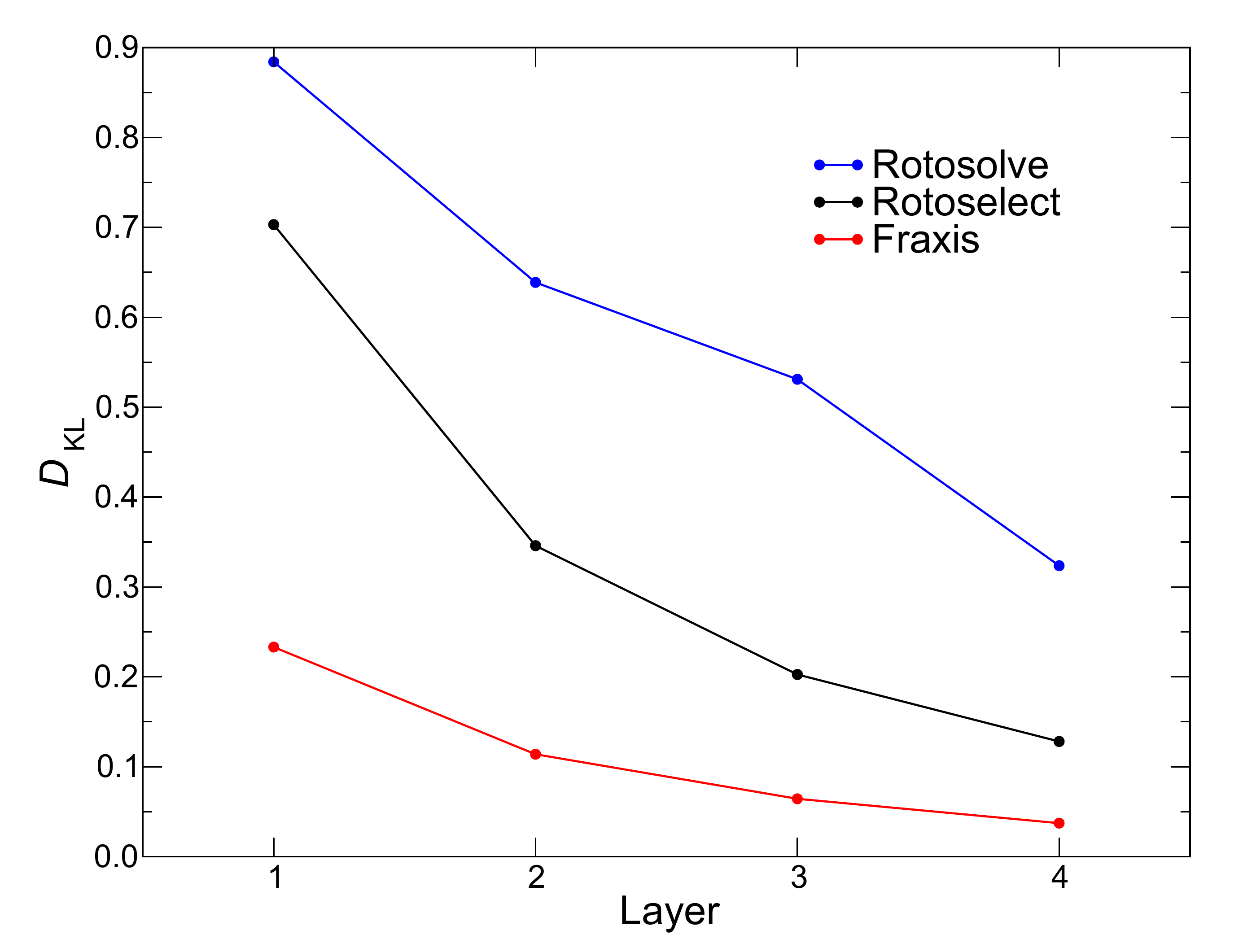} \\
  \end{tabular}
  \caption{Relation between KL divergences (lower is better) and number of layers for circuits A (left) and B (right)}
  \label{fgr:KLdivergence}
\end{figure*}

Next, we evaluated expressibility of two multi-layer PQCs as shown in Fig.~\ref{fgr:circuits} consisting of 5 and 6 qubits respectively. Fig.~\ref{fgr:KLdivergence} shows the relation between the KL divergence and number of layers. 
It is notable that Fraxis always obtained smaller KL divergences than Rotosolve and Rotoselect.

\subsection{Heisenberg Model}

We performed energy optimization using the Hamiltonian that is one dimensional Heisenberg model under the periodic boundary condition with 5-qubit as
\begin{equation}
\label{Heisenberg}
    H= J \sum_{(i,j)\in\mathcal{E}}(X_i X_j +Y_i Y_j + Z_i Z_j ) + h\sum_{i\in \mathcal{V}} Z_i,
\end{equation}
where $J=h=1$. We employed circuit A in Fig.~\ref{fgr:circuits}, in Rotosolve, Rotoselect and Fraxis ansatze, respectively. In Rotosolve, all gates were represented by $R_y$. Although the gate update order is arbitrary and can affect the final results, we here conducted gate update as: starting from the first layer, ($i$) update the single-qubit gate at the $0$-th qubit in the layer, ($ii$) update each single-qubit gate in the ascending order of qubits in the layer, ($iii$) move to the next layer in ascending order, and ($iv$) go back to ($i$) from the first layer until maximum number of iterations is reached.   

\begin{figure*}[tb]
  \centering
  \begin{tabular}{cc}
  \includegraphics[scale=0.25]{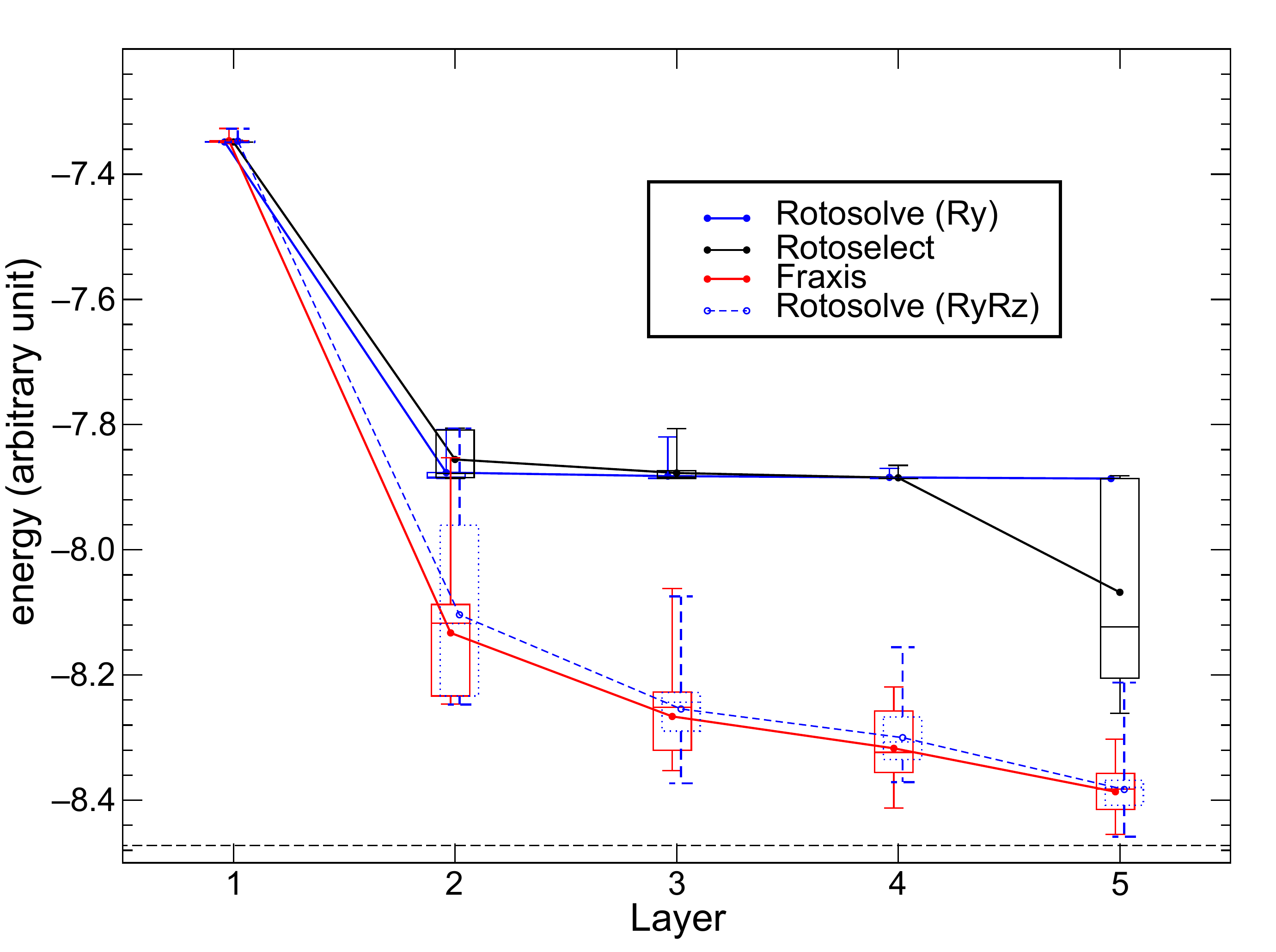} & 
  \includegraphics[scale=0.25]{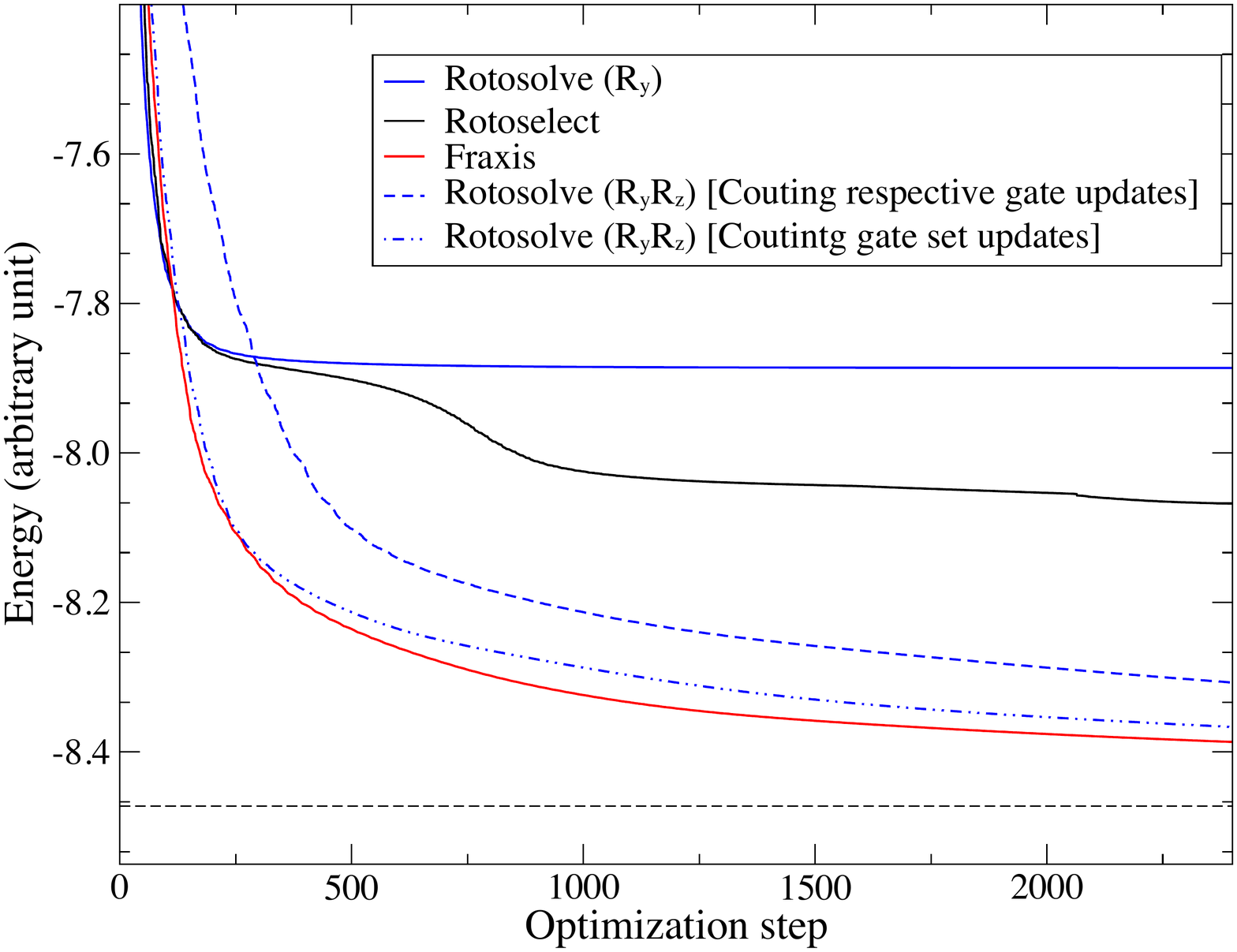} \\
  \end{tabular}
  \caption{(Left) Relation between number of layers for circuit A and the optimized energy after 100 iterations for Rotosolve, Rotoselect, and Fraxis. The averaged energy is represented by filled circles and lines. The boxplots represent respective quantiles. The exact ground state energy is displayed with dash line. (Right) The achieved energy against the number of optimization steps for Rotosolve, Rotoselect, and Fraxis. At both subfigures, the Rotosolve is also applied to $R_yR_z$ gates whose degrees of freedom are the same as Fraxis.}
  \label{fgr:opt_heisenberg}
\end{figure*}

Figure~\ref{fgr:opt_heisenberg} shows the optimized energy after 100 iteration and thus the optimization is not necessarily converged.
Although KL divergence shows large difference among the ansatze in Fig.~\ref{fgr:KLdivergence}, the obtained energy are consistent among all ansatze for $L=1$.
In the multi-layer calculation with the Rotosolve ansatz using $R_y$ gates, the energy were not improved except for at $L=2$. 
It is probably because the ansatz consisting of only $R_y$ gates cannot cover the ground state of this model. For this reason, we also run the Rotosolve ansatz using $R_yR_z$ gates and confirm that the obtained energies were much improved but still slightly inferior to Fraxis as shown in both subfigures. The figures present empirical evidence that Fraxis is still better than the Rotosolve having the same degree of freedom with $R_yR_z$ gates.

It is also notable that although the expressibility of Rotoselect was not improved by increasing the number of layers from 1 to 2, the obtained energy was obviously improved in Rotosolve. 
Fraxis showed significant improvement even at $L=$ 3, 4, and 5, although the improvement of the expressibility was not as large as those of Rotosolve and Rotoselect in Fig.~\ref{fgr:KLdivergence}.
Altogether, although Fraxis showed the best performance, it seems that its performance cannot be explained solely by the expressibility measured with KL divergence.
And thus, it is also important to discuss the circuit performance with the entanglement capability.
However, it should be stressed that the Fraxis ansatz always showed the best performance with the given circuit and the number of layers.

\subsection{Molecular Hamiltonian Model: \ce{H2}/6-31G}
\begin{figure}[tb]
  \centering
  \includegraphics[scale=0.3]{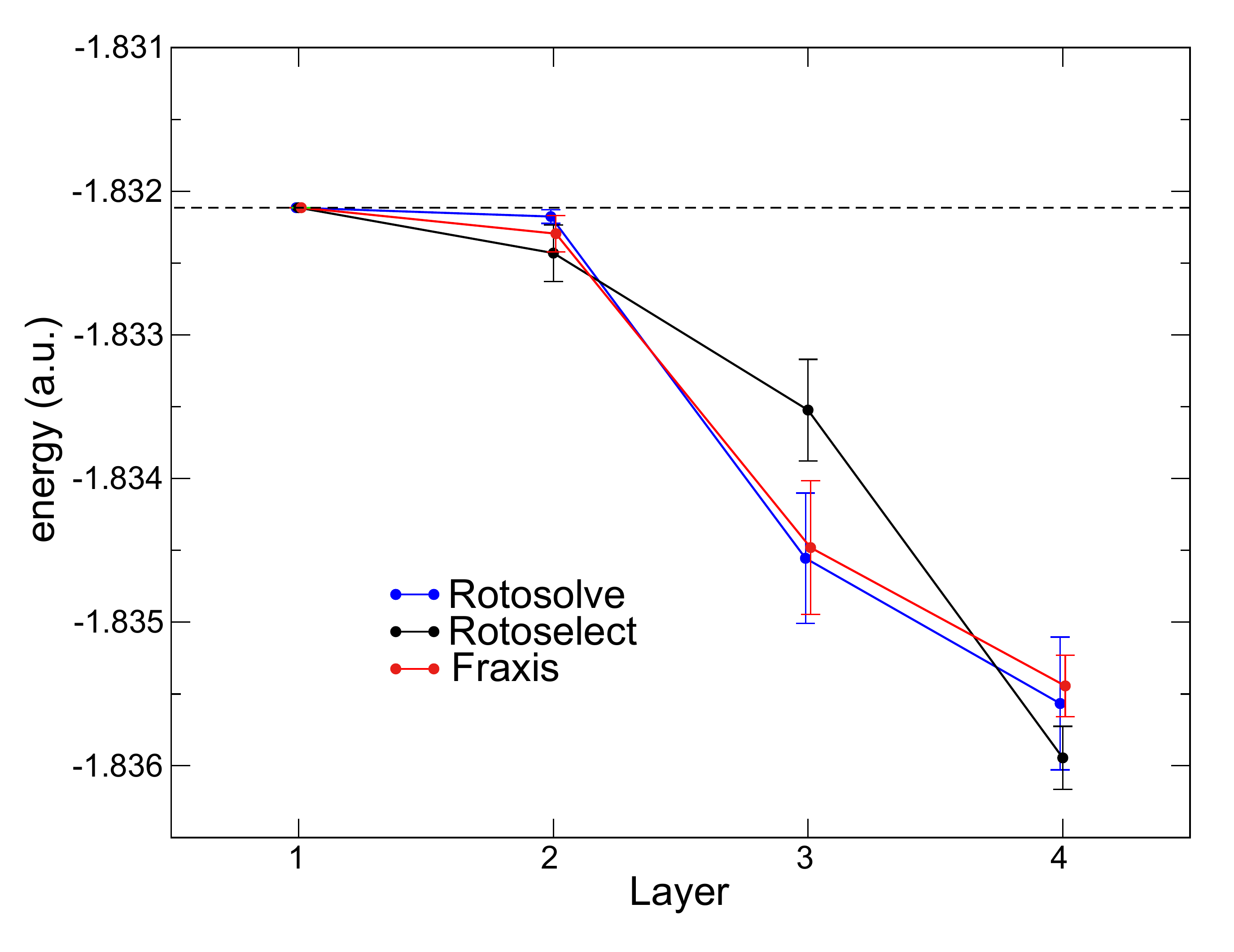}
  \caption{The average of optimized energy after 100 iterations with standard error. The Hartree-Fock state energy is displayed with dash line.}
  \label{fgr:opt_H2}
\end{figure}
Here, we evaluated the electronic energy of \ce{H2} molecules with bond length of 0.75 \AA.
First we performed the Hartree-Fock calculation with 6-31G basis set, which was followed by the fermionic Hamiltonian construction with active space [2,3].
Then, the fermionic Hamiltonian was converted to the qubit Hamiltonian by Jordan-Wigner transformation.
The optimization continued until either reaching to convergence or 100 iterations, where respective gates were sequentially updated as in the previous section. 
Since the optimization strongly depends on the initial conditions, we evaluated statistical average by randomly generating the initial parameters. 
Paying attention to the statistical error, the independent optimizations were performed 20 to 300 times with different initial conditions.
In the case of Fraxis, the initial states were generated in the manner of parameter random as described in Sec.~\ref{subsec:circ-exp}.

In agreement with the result of the Heisenberg model, the obtained energy did not vary among ansatze at $L=1, 2$ regardless of the diverse expressibility and all optimizations were stuck in local minimum at the Hartree-Fock energy level in Fig.~\ref{fgr:opt_H2}. Note that at $L=3$ the averaged optimized energy by Rotosolve is lower than that by Rotoselect.
For qualitative understanding of this non-intuitive results,
it should be noted that in quantum chemistry, the coefficients of the CI expansion are in usual real values, which indicates that the eigenstates can be expressed by only $R_y$ gates. 
We suppose the high expressibility can bring two conflicting consequences, (1) inclusion of the crucial piece of Hilbert space spanned by the ansatz, which may lead to successfully the ground state. This also includes the ability to express the ground state itself, (2) inessential extension of search space irrelevant to the efficient optimization.
We suppose that the latter effects emerged in the case of Rotoselect at $L=3$ because the Rotoselect ansatz greedily selects a gate that yields the lowest energy at each optimization state.
Indeed, we confirmed that Rotoselect optimization did not necessarily select $R_y$ gate, which may not be the best choice to find the global minimum, although it can temporally gain the largest energy decrease.
In contrast, Fraxis resulted in the energy at the same level as Rotosolve.
Although we found the selected axes in Fraxis were not necessarily on the X-Z plain of the Bloch sphere expression, it may be possible to express effective path to the ground state by utilizing complex space if the expresssibility is sufficiently high.
As supporting evidences, the optimized efficiency levels of Rotoselect and Fraxis are reversed at $L=4$, which is probably because as the number of layer increases, the confined search space in Rotosolve and Rotoselect may help narrowing the search space. 

\subsection{Hamiltonian for MaxCut}
We want to show the applicability of the Fraxis gates when the optimal quantum state is not spanned by quantum states with real amplitudes, and when its approximation is sufficient (i.e., its exact quantum state is not required). A quantum relaxation for combinatorial optimization in~\cite{fuller2021approximate} is one of the methods that rely on finding such quantum states to solve MaxCut problems. Given a graph $G=(V,E)$ with $|V|$ vertices and $|E|$ edges, the MaxCut problem, which is NP-hard, asks for assigning each vertex in the graph to either $1$ or $-1$ so that the sum of edges connecting vertices with different values is maximized. Namely, it finds a configuration $m\in\{-1,1\}^{|V|}$ that maximizes a cost function ${\rm cut}(m)$, where 
\begin{equation}
    {\rm cut}(m) 
    \equiv \sum_{(i,j)\in E}\frac{1}{2}(1-m_i m_j). 
\end{equation}

\usetikzlibrary{graphs,graphs.standard}
\tikzgraphsset{edges={draw,semithick}, nodes={circle,draw,semithick}}

\begin{figure*} [tb]
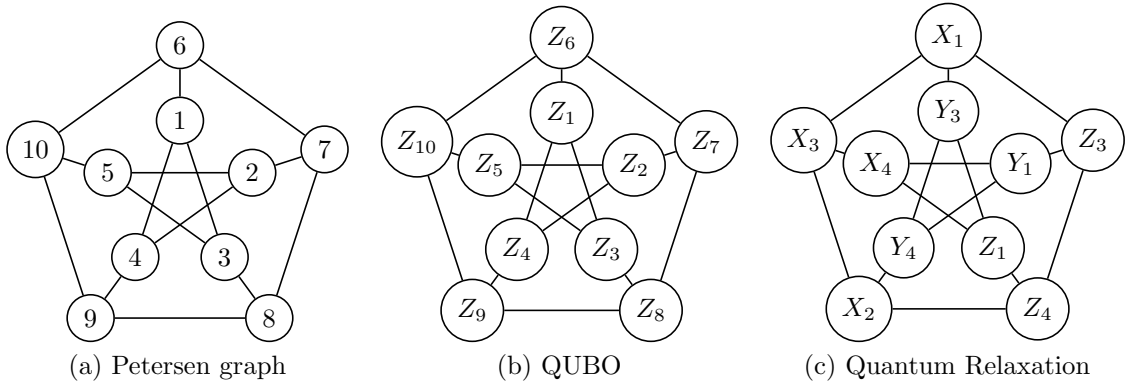

\centering
\begin{tabular}{ccc}
\tikz \graph[math nodes, clockwise]
    { subgraph I_n [V={1,2,3,4,5}] --
      subgraph C_n [V={6,7,8,9,10},radius=1.00cm];
      {[cycle] 1,3,5,2,4} };&
\tikz \graph[math nodes, clockwise]
    { subgraph I_n [V={Z_1,Z_2,Z_3,Z_4,Z_5}] --
      subgraph C_n [V={Z_6,Z_7,Z_8,Z_9,Z_{10}},radius=1.00cm];
      {[cycle] Z_1,Z_3,Z_5,Z_2,Z_4} };&
\tikz \graph[math nodes, clockwise]
    { subgraph I_n [V={Y_3,Y_1,Z_1,Y_4,X_4}] --
      subgraph C_n [V={X_1,Z_3,Z_4,X_2,X_3},radius=1.00cm];
      {[cycle] Y_3,Z_1,X_4,Y_1,Y_4} };\\
      (a) Petersen graph & 
      (b) QUBO & 
      (c) Quantum Relaxation\\
\end{tabular}
\caption{The Petersen graph (Left), the labelling of its vertices to construct the Hamiltonian for QUBO (Center), and the corresponding labelling to construct the Hamiltonian for Quantum Relaxation (Right) whose optimal quantum state can be highly entangled with non real-valued amplitudes.}\label{fig:petersen}
\end{figure*}

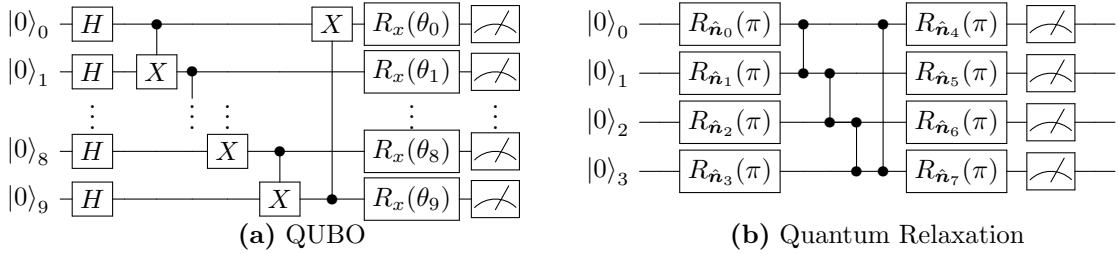
\begin{figure*} [tb]
\centering
\begin{tabular}{cccccccccc}

\Qcircuit @C=.45em @R=.2em {
  \lstickx{\ket{0}_0} & \gate{H}& \qw & \ctrl{1}          & \qw              & \qw& \qw  & \gate{X} & \gate{R_x(\theta_{0})}  & \meter \\
  \lstickx{\ket{0}_1} & \gate{H}& \qw & \gate{X}    &\ctrl{1}          & \qw& \qw               & \qw  &  \gate{R_x(\theta_{1})}   & \meter  \\
  &&&  &          &           & &  & &   &    & \\
  &&&  &          &           & &  & &   &    & \\
  &  \vdots  &     &        &  \vdots          &  \vdots             &                      &        &       \vdots    &        \vdots  &            &      \\
  &&&  &          &           & &  & &   &    & \\
  &&&  &          &           & &  & &   &    & \\
  &&&  &          &           & &  & &   &    & \\
  \lstickx{\ket{0}_8} & \gate{H}& \qw  &\qw &\qw & \gate{X}     &\ctrl{1}            & \qw      &  \gate{R_x(\theta_{8})} & \meter \\
  \lstickx{\ket{0}_9} & \gate{H}& \qw  &\qw &\qw & \qw               & \gate{X}     & \ctrl{-9}      & \gate{R_x(\theta_{9})}  & \meter    \\
}
&~&~&
\Qcircuit @C=.75em @R=.25em {
  \lstickx{\ket{0}_0} & \qw & \gate{\UU[0]} & \ctrl{1}          & \qw              & \qw& \control \qw& \gate{\UU[4]}  & \meter   & \qw     & \qw \\
  \lstickx{\ket{0}_1} & \qw & \gate{\UU[1]} & \control \qw & \ctrl{1}          & \qw& \qw& \gate{\UU[5]}  & \meter    & \qw      & \qw \\
  \lstickx{\ket{0}_2} & \qw & \gate{\UU[2]} & \qw         & \control \qw  & \ctrl{1} \qw & \qw& \gate{\UU[6]}  & \meter    & \qw     & \qw \\
  \lstickx{\ket{0}_3} & \qw & \gate{\UU[3]} &  \qw & \qw       &  \control \qw & \ctrl{-3}& \gate{\UU[7]} & \meter   & \qw      & \qw \\
}\\
\textbf{(a)} QUBO 
&~&~&
\textbf{(b)} Quantum Relaxation
\end{tabular}
\caption{PQCs for QUBO and Quantum Relaxation of the Petersen graph}\label{fgr:maxcut_circuit}
\end{figure*}

Figure~\ref{fig:petersen} shows a particular example of MaxCut instances considered in~\cite{fuller2021approximate}. The left subfigure shows the Petersen graph that consists of $10$ nodes and $15$ edges. The center subfigure illustrates the Quadratic Unconstrained Binary Optimization (QUBO) formulation of the MaxCut problem of the Petersen graph, where the Hamiltonian of the QUBO formulation is defined as 
\begin{equation}
    H_{\mbox{QUBO}} \equiv \sum_{(i,j)\in E}\frac{1}{2}(1-Z_i Z_j),
\end{equation}
where $Z_i Z_j$ denotes the tensor product of Pauli matrices $Z$ at the $i$-th and $j$-th qubit, and $I$ at the rest of the qubits. Popular quantum optimization methods, such as, QAOA~\cite{QAOA2014}, utilize the QUBO formulation to encode a binary variable into a qubit. Namely, the QUBO formulates a $10$-qubit Hamiltonian to solve the MaxCut problem of the Petersen graph to obtain an optimal $10$-qubit quantum state.
On the other hand, the Quantum Relaxation utilizes an encoding that can map three binary variables into a qubit. We refer the readers to~\cite{fuller2021approximate} for more details on the mapping and the resulting Hamiltonian. For the Petersen graph, the Hamiltonian of the Quantum Relaxation is given as 
\begin{equation}
    H_{\mbox{Relax}} \equiv \sum_{(i,j)\in E}\frac{1}{2}(1-3~W_i W_j),
\end{equation}
where $W_i \in \{X_1, Y_1, Z_1, \ldots, Z_4\}$ is the Pauli matrix that can be inferred from the label of the vertex as shown in the center subfigure of Fig.~\ref{fig:petersen}. Here, we can observe that the $H_{\mbox{QUBO}}$ is a diagonal matrix (because it consists of the weighted sum of tensor products of Pauli $I$ and $Z$), while the $H_{\mbox{Relax}}$ is not (because it consists of the weighted sum of Pauli $I, X, Y$ and $Z$). Consequently, it is easy to see that the optimal quantum state of the former is a computational basis, while that of the latter is not necessarily so. The striking difference is in the size of the Hamiltonian: $2^{10}\times 2^{10}$ for the former and $2^4 \times 2^4$ for the latter. Because of its smaller Hamiltonian, the Quantum Relaxation requires smaller PQCs to find an optimal quantum state of the underlying problem.

One of major obstacles is how to recover the binary solution from the optimal quantum state $\rho^*$, which is unlike that of QUBO, can be highly entangled in non computational basis. A simple \textit{magic} rounding method that maps $\rho^*$ into a random quantum state which is a product of a one-qubit state in the form of $ \frac{1}{2}\left(I + \frac{1}{\sqrt{3}}\left(\pm X \pm Y \pm Z\right)\right)$ (such states are called \textit{magic states} and hence the method name) is shown to guarantee the solution of MaxCut whose cut value is $5/9$ of the optimal one. Moreover, such guarantee holds not only for the optimal $\rho^*$, but also for any quantum state $\rho_{\mbox{relax}}$ satisfying $\mathrm{Tr}(H_{\mbox{Relax}}\rho_{\mbox{relax}}) \ge {\rm cut}(m) $. Namely, it suffices to find a quantum state whose relaxed Hamiltonian value exceeds that of the optimal binary solution. The quality of the rounding from such quantum state is determined by the Hamiltonian value: the higher it is the better. Notice that although the approximation guarantee $5/9$ is lower than the best of classical one~\cite{GW1995}, in practice the MaxCut solution of Quantum Relaxation can be much better. See ~\cite{fuller2021approximate} for more details.

We provide evidences that PQCs with Fraxis gates can obtain better quantum states for Hamitonians of QUBO and Quantum Relaxation. We followed the construction of Hamiltonians as in~\cite{fuller2021approximate} for the Petersen graph, as shown in Fig.~\ref{fig:petersen} and test PQCs with single-qubit parametrized gates, as shown in Fig.~\ref{fgr:maxcut_circuit}, to obtain $\rho_{\mbox{relax}}$.  The PQCs on the left of Fig.~\ref{fgr:maxcut_circuit} were used for QUBO Hamiltonian with the Hadamard gates in the first layer of single-qubit gates so that the initial states is the superposition of all possible cuts as suggested in~\cite{QAOA2014}. The entangling gates are fixed to cyclic CX gates with single-qubit gates varied from fixed axes to Fraxis (the figure depicts the fixed axes ones). The PQCs on the right were for the Quantum Relaxation Hamiltonian with cyclic CZ gates as entanglers and the single-qubit gates varied similarly. 

\begin{figure}[tb]
  \centering
  \begin{tabular}{cc}
  \includegraphics[scale=0.25]{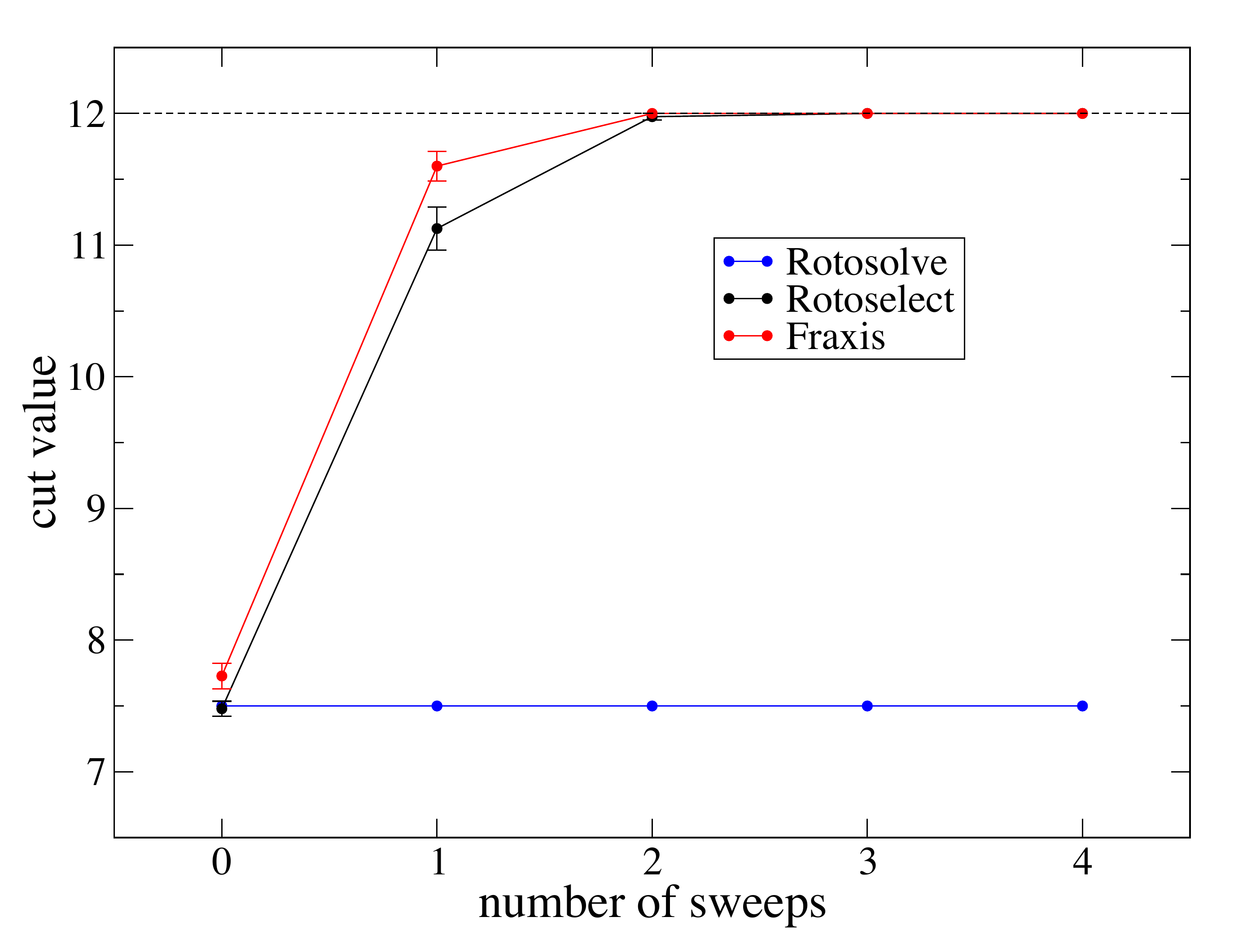} &
  \includegraphics[scale=0.25]{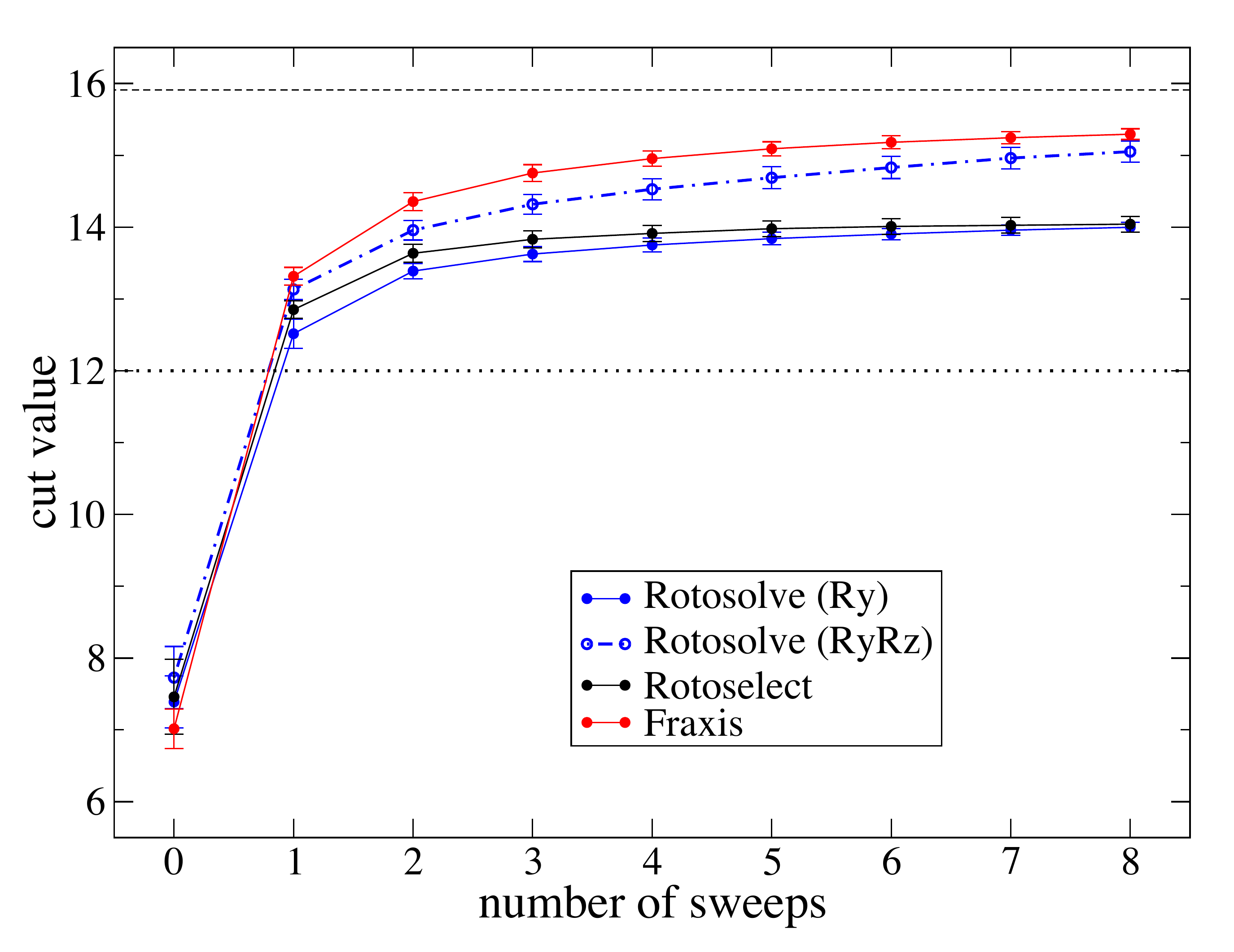}\\
  (a) QUBO &
  (b) Quantum Relaxation
   \end{tabular}
   \caption{The cut values (higher is better) of quantum states against the number of sweeps updating parameters of PQCs with different single-qubit gates for QUBO (Top) and Quantum Relaxation (Bottom) averaged over 20 independent trajectories for the Maxcut on the Petersen graph. The dash and dotted lines represent the optimal quantum relaxed value and the optimal binary solution, resp. For both QUBO and Quantum Relaxation, PQCs with Fraxis gates find better quantum states.}  
  \label{fgr:maxcut_trajectories}
\end{figure}

First, we carried out 20 independent optimizations from $H^{\otimes 10}\left|0\right> $ for the QUBO Hamiltonian using PQCs with fixed-axis single-qubit gates optimized with Rotosolve and Rotoselect against those with Fraxis. The top figure in Fig.~\ref{fgr:maxcut_trajectories} shows the plots of the expected cut values against the number of parameter updates of the single-qubit gates of the PQCs. We observe that fixing the single-qubit gates to $R_y$ or $R_x$ and updating their parameters with Rotosolve fails as the cut values do not improve from the initial value. This is surprising as PQCs with $R_y$ or $R_x$ gates are often suggested, such as in~\cite{Moll2018}, for the Hamiltonian whose ground states are of real-valued amplitudes. On the other hand, Rotoselect and Fraxis successfully find the optimal cuts after two sweeps of parameter updates despite their larger search space of complex-valued amplitudes. 

Next, we carried out 20 independent optimizations from different initial conditions using the ansatz with cyclic entangler to find the ground state of the Quantum Relaxation Hamiltonian using PQCs with Rotosolve, Rotoselect and Fraxis. Unlike the QUBO and the quantum chemistry Hamiltonians, the eigenstate of the Quantum Relaxation Hamiltonian cannot be generated by the $R_y$ ansatz alone, and thus, $R_y$ ansatz may be unfairly disadvantageous. 
Hence, we independently carried out two types of Rotosolve optimizations with both $R_y$ and $R_yR_z$ ans\"{a}tze. We confirm that Rotosolve on PQCs with $R_yR_z$ ansatz are better than that on PQCs with only $R_y$ ansatz, or even with fixed-axis single-qubit gates optimized with Rotoselect (namely, parametrized gates chosen from $R_x$, $R_y$, and $R_z$ rotations). Here, we can see that PQCs with Fraxis obtain better cut values when compared to the rest of PQCs. Hence, freeing the design of PQCs on the choices of single-qubit gates is not only convenient but also can result in better optimization values.

\subsection{Experimental Evaluation on Near-Term Quantum Devices}
We compared the performance of Rotosolve and Fraxis using two types of 2-qubit Hamiltonian on the 5-qubit \textsf{ibmq\_manila} quantum device of the IBM Quantum Systems. The experimental evaluations were performed on the device from May 22 to May 24, 2021 by \textit{fairshare} run mode. 
The employed PQCs were homologous of that in Fig.~\ref{fgr:2qubit_PQC} but all $\theta$-Fraxis gates were replaced by $\pi$-Fraxis gates.

\begin{figure}[tb]
  \centering
  \begin{tabular}{cc}
  \includegraphics[scale=0.25]{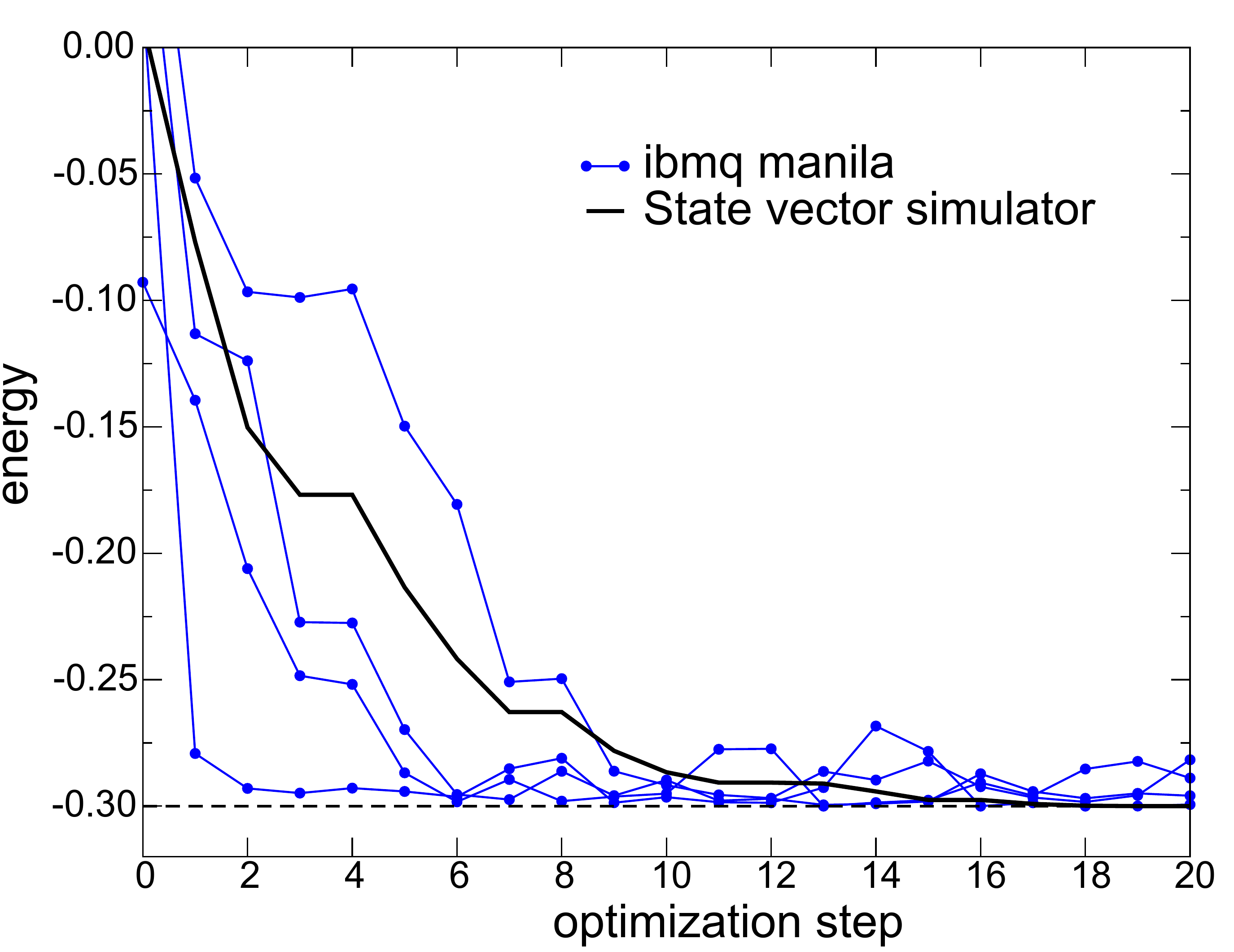} & 
  \includegraphics[scale=0.25]{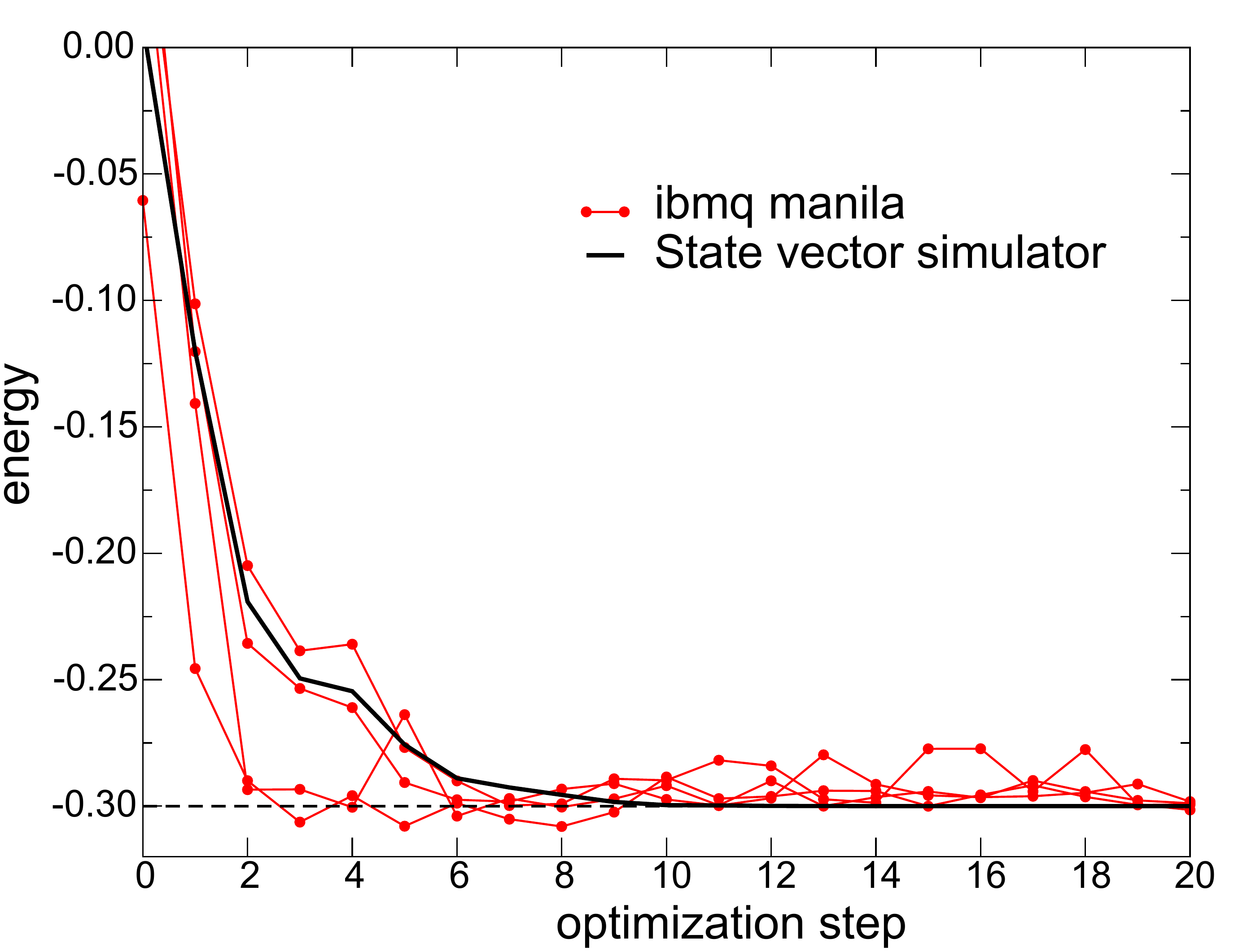}\\
  (a) Rotosolve &
  (b) Fraxis\\
   \end{tabular}
   \caption{Comparison of energy convergence as a number of optimization steps for Heisenberg model on the \textsf{ibmq\_manila} device with readout error mitigation.
   (a) Rotosolve (blue), (b)Fraxis (red).
   Both method are applied to start from the 4 random initial parameter sets.}  
  \label{fgr:device_heisenberg}
\end{figure}

First, we calculated the 2-qubit Heisenberg model which is given in~\eqref{2qubit-model}.
Figure~\ref{fgr:device_heisenberg} shows four energy trajectories for Rotosolve and Fraxis, respectively, where the initial parameter sets were randomly generated.
Here, the energy expectation values were evaluated from 8192 measurements (or, shots) with readout error mitigation.
Although, according to the simulator result in Fig.~\ref{fgr:comparison}, Rotosolve can be trapped in local minima, it seemed to converge to the ground state in all four trials.
However, its convergence seemed to largely vary among the trials.
In contrast, Fraxis showed better convergence (e.g., less number of optimization steps to reach the ground state in all four trials) compared to Rotosolve in consistency with the simulator results in Sec.~\ref{subsec:com_opt}.

\begin{figure}[tb]
  \centering
  \begin{tabular}{cc}
  \includegraphics[scale=0.25]{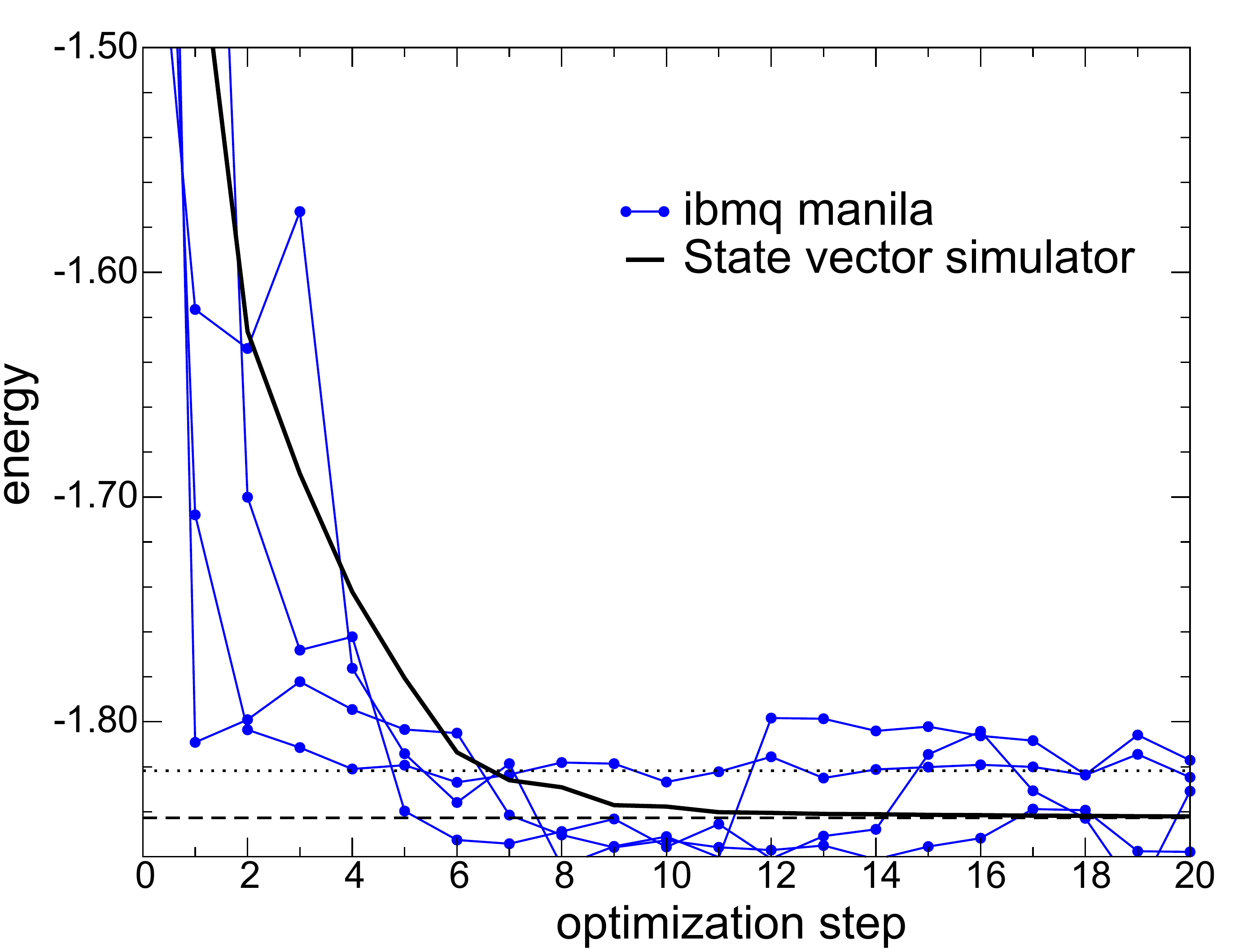} & 
  \includegraphics[scale=0.25]{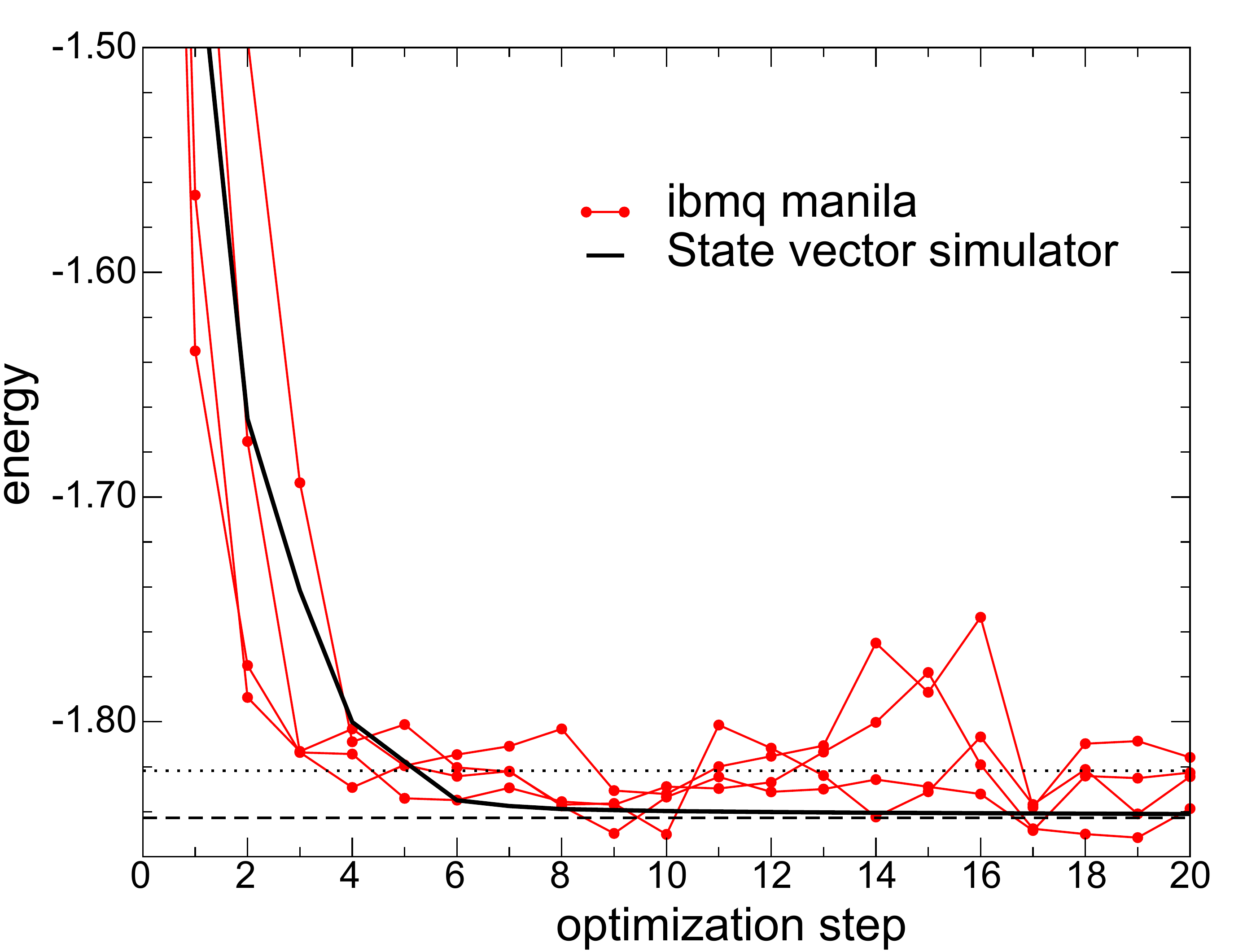}\\
  (a) Rotosolve &
  (b) Fraxis
   \end{tabular}
   \caption{Comparison of energy convergence as a number of optimization steps for \ce{H2}/STO-3G on the \textsf{ibmq\_manila} device with readout error mitigation.
   (a) Rotosolve (blue), (b)Fraxis (red).
   Both method are applied to start from the 4 random initial parameter sets.
   The exact ground state energy is displayed with dashed line.
   The Hartree-Fock state is displayed with dotted line.} 
  \label{fgr:device_H2sto3g}
\end{figure}

Next, we also benchmarked the \ce{H2} molecular Hamiltonian with STO-3G basis set as shown in Fig.~\ref{fgr:device_H2sto3g}. 
In an ideal case, the energy should monotonically decreases in both Rotosolve and Fraxis in course of optimization step as shown with the simulator in Sec.~\ref{subsec:com_opt}, although it was not the case.
Such unwarranted behavior did not vary between Rotosolve and Fraxis.
However, it seemed to be more distinct with \ce{H2} Hamiltonian rather than Heisenberg model, which may reflect difference in number of Pauli terms and corresponding accumulated errors.
As a result, it is difficult to obtain the converged value of \ce{H2} under Hartree-Fock energy.
Thus, for effective utilization of NISQ in chemistry, it may be necessary to utilize other error correction such as state tomography as demonstrated in literature\cite{gao2021applications}.
Since PQCs consisting of $R_y$ gates are sufficient to express the ground state, as discussed before, Rotosolve may be more advantageous in converging efficiency because of its confined search space.
Nevertheless we did not find any distinct difference between Rotosolve and Fraxis. In fact, one of the four trials of Rotosolve seemed to be trapped in the Hartree-Fock energy up to 20 optimization steps, while all four trials of Fraxis could find energies lower than the Hartree-Fock energy within the same steps. Although the search space of Fraxis is not confined within the real-valued probability amplitudes, Fraxis can still find the ground state effectively. 

Overall, we can observe from Figs.~\ref{fgr:device_heisenberg} and~\ref{fgr:device_H2sto3g} that on 2-qubit systems the performance of Fraxis (and Rotosolve) with real quantum devices is similar to that with noiseless simulators. This hints the potential usefulness of Fraxis-ansatz for a variety of Hamiltonians with near-term quantum devices.

%% file: conclusion.tex
In this paper, we proposed a method of free-axis selection optimizing rotational axes of single-qubit gates in PQCs.
The Fraxis ansatz can achieve high expressibility even with shallow PQCs. We confirmed the expressibilities estimated with KL divergence for both single- and multi-qubit PQCs with Fraxis are significantly higher when compared to optimizing the rotation angles. Nevertheless, only six energy estimations are required for each gate update to find the optimal axis, which is more efficient than Rotoselect that requires seven of them. Fraxis ansatz seems to be more effective in dealing with general quantum systems, such as, the Heisenberg model whose ground states may not be real-valued amplitudes like those of molecular Hamiltonian models. Meanwhile, \cite{govia2021freedom} recently showed that a restricted Fraxis ansatz can improve QAOA for MaxCut despite the ground states are real-valued amplitudes. We confirmed numerically that Fraxis ansatz could be better at similar cases to find ground states of Hamiltonian for MaxCut. 

Although Fraxis enables to find the optimal rotation axis for single qubit gate, the optimal order of gate update still remains untouched. Indeed, as seen in the application to a \ce{H2} molecules, many optimization runs were trapped in Hartree--Fock states, which also happened with Rotosolve and Rotoselect.
This result implies the optimal choice for a single qubit rotation is not necessarily the best strategy for the entire system to avoid local minima and Barren plateaus. This may be avoided by randomizing the order of updates and/or by gradient-based approaches to avoid local minima, which we plan to pursue as future work. Another promising approach is by simultaneously updating $k$ axes, which is similarly shown in~\cite{Nakanishi2020PRR,Parrish2019arXiv} to update $k$ angles simultaneously. However, updating $k$ axes by Fraxis requires $6^k$ energy estimations in contrast to $3^k$ energy estimation of structural optimization in~\cite{Nakanishi2020PRR,Parrish2019arXiv} and for this reason we may have to combine Fraxis with other techniques, such as, matrix completion~\cite{Candes2009,CandesTao2010}. Fraxis' higher number of energy estimations for classical optimization of parameters may be unavoidable due to its higher expressibility. 
Thanks to the matrix factorization framework of Fraxis, sequential arbitrary single-qubit optimization~\cite{WRSW2022} and sequential time evolution of quantum systems in~\cite{WROKSYW2022} have been successfully developed. The codes of Fraxis and its extension are publicly available.

%% file: main.bbl
\begin{thebibliography}{10}
\providecommand{\url}[1]{#1}
\csname url@samestyle\endcsname
\providecommand{\newblock}{\relax}
\providecommand{\bibinfo}[2]{#2}
\providecommand{\BIBentrySTDinterwordspacing}{\spaceskip=0pt\relax}
\providecommand{\BIBentryALTinterwordstretchfactor}{4}
\providecommand{\BIBentryALTinterwordspacing}{\spaceskip=\fontdimen2\font plus
\BIBentryALTinterwordstretchfactor\fontdimen3\font minus
  \fontdimen4\font\relax}
\providecommand{\BIBforeignlanguage}[2]{{%
\expandafter\ifx\csname l@#1\endcsname\relax
\typeout{** WARNING: IEEEtran.bst: No hyphenation pattern has been}%
\typeout{** loaded for the language `#1'. Using the pattern for}%
\typeout{** the default language instead.}%
\else
\language=\csname l@#1\endcsname
\fi
#2}}
\providecommand{\BIBdecl}{\relax}
\BIBdecl

\bibitem{Cerezo2020arXiv}
M.~Cerezo~\textit{et al}, ``Variational {Q}uantum {A}lgorithms,''
  \emph{arXiv:2012.09265}, 2020.

\bibitem{Bharti2021arXiv}
K.~Bharti~\textit{et al}, ``Noisy intermediate-scale quantum ({NISQ})
  algorithms,'' \emph{arXiv:2101.08448}, 2021.

\bibitem{TillyetalVQE2021}
\BIBentryALTinterwordspacing
J.~Tilly, H.~Chen, S.~Cao, D.~Picozzi, K.~Setia, Y.~Li, E.~Grant, L.~Wossnig,
  I.~Rungger, G.~H. Booth, and J.~Tennyson, ``The variational quantum
  eigensolver: a review of methods and best practices,'' 2021. [Online].
  Available: \url{https://arxiv.org/abs/2111.05176}
\BIBentrySTDinterwordspacing

\bibitem{Peruzzo2014NatCom}
\BIBentryALTinterwordspacing
A.~Peruzzo~\textit{et al}, ``A variational eigenvalue solver on a photonic
  quantum processor,'' \emph{Nat. Commun.}, vol.~5, no.~1, p. 4213, 2014.
  [Online]. Available: \url{https://doi.org/10.1038/ncomms5213}
\BIBentrySTDinterwordspacing

\bibitem{QAOA2014}
\BIBentryALTinterwordspacing
E.~Farhi, J.~Goldstone, and S.~Gutmann, ``A quantum approximate optimization
  algorithm,'' 2014. [Online]. Available: \url{https://arxiv.org/abs/1411.4028}
\BIBentrySTDinterwordspacing

\bibitem{Preskill2018quantumcomputingin}
\BIBentryALTinterwordspacing
J.~Preskill, ``Quantum {C}omputing in the {NISQ} era and beyond,''
  \emph{{Quantum}}, vol.~2, p.~79, Aug. 2018. [Online]. Available:
  \url{https://doi.org/10.22331/q-2018-08-06-79}
\BIBentrySTDinterwordspacing

\bibitem{Kandala2017Nat}
A.~Kandala~\textit{et al}, ``Hardware-efficient variational quantum eigensolver
  for small molecules and quantum magnets,'' \emph{Nature}, vol. 549, pp.
  242--246, 2017.

\bibitem{Barkoutsos2018PRA}
\BIBentryALTinterwordspacing
P.~K. Barkoutsos~\textit{et al}, ``Quantum algorithms for electronic structure
  calculations: Particle-hole {H}amiltonian and optimized wave-function
  expansions,'' \emph{Phys. Rev. A}, vol.~98, p. 022322, Aug 2018. [Online].
  Available: \url{https://link.aps.org/doi/10.1103/PhysRevA.98.022322}
\BIBentrySTDinterwordspacing

\bibitem{Ganzhorn2019PRApp}
\BIBentryALTinterwordspacing
M.~Ganzhorn~\textit{et al}, ``Gate-{E}fficient {S}imulation of {M}olecular
  {E}igenstates on a {Q}uantum {C}omputer,'' \emph{Phys. Rev. Applied},
  vol.~11, p. 044092, Apr 2019. [Online]. Available:
  \url{https://link.aps.org/doi/10.1103/PhysRevApplied.11.044092}
\BIBentrySTDinterwordspacing

\bibitem{Gard2020npjQI}
\BIBentryALTinterwordspacing
B.~T. Gard, L.~Zhu, G.~S. Barron, N.~J. Mayhall, S.~E. Economou, and E.~Barnes,
  ``Efficient symmetry-preserving state preparation circuits for the
  variational quantum eigensolver algorithm,'' \emph{npj Quantum Inf.}, vol.~6,
  no.~1, p.~10, 2020. [Online]. Available:
  \url{https://doi.org/10.1038/s41534-019-0240-1}
\BIBentrySTDinterwordspacing

\bibitem{Tang2021PRXQ}
\BIBentryALTinterwordspacing
H.~L. Tang~\textit{et al}, ``Qubit-{ADAPT}-{VQE}: An {A}daptive {A}lgorithm for
  {C}onstructing {H}ardware-{E}fficient {A}ns\"atze on a {Q}uantum
  {P}rocessor,'' \emph{PRX Quantum}, vol.~2, p. 020310, Apr. 2021. [Online].
  Available: \url{https://link.aps.org/doi/10.1103/PRXQuantum.2.020310}
\BIBentrySTDinterwordspacing

\bibitem{Tkachenko2020arXiv}
N.~V. Tkachenko~\textit{et al}, ``{Correlation-Informed Permutation of Qubits
  for Reducing Ansatz Depth in VQE},'' \emph{arXiv:2009.04996}, 2020.

\bibitem{McClean2018NatComm}
J.~R. McClean, S.~Boixo, V.~N. Smelyanskiy, R.~Babbush, and H.~Neven, ``Barren
  plateaus in quantum neural network training landscapes,'' \emph{Nat.
  Commun.}, vol.~9, 2018.

\bibitem{Cerezo2021NatComm}
M.~Cerezo, A.~Sone, T.~Volkoff, L.~Cincio, and P.~J. Coles, ``Cost function
  dependent barren plateaus in shallow parametrized quantum circuits,''
  \emph{Nat. Commun.}, vol.~12, 2021.

\bibitem{Pesah2020arXiv}
A.~Pesah, M.~Cerezo, S.~Wang, T.~Volkoff, A.~T. Sornborger, and P.~J. Coles,
  ``Absence of barren plateaus in quantum convolutional neural networks,''
  \emph{arXiv:2011.02966}, 2020.

\bibitem{Holmes2021PRL}
\BIBentryALTinterwordspacing
Z.~Holmes, A.~Arrasmith, B.~Yan, P.~J. Coles, A.~Albrecht, and A.~T.
  Sornborger, ``Barren plateaus preclude learning scramblers,'' \emph{Phys.
  Rev. Lett.}, vol. 126, p. 190501, May 2021. [Online]. Available:
  \url{https://link.aps.org/doi/10.1103/PhysRevLett.126.190501}
\BIBentrySTDinterwordspacing

\bibitem{Zhao2021arXiv}
C.~Zhao and X.-S. Gao, ``Analyzing the barren plateau phenomenon in training
  quantum neural network with the {ZX}-calculus,'' \emph{arXiv:2102.01828},
  2021.

\bibitem{Arrasmith2020arXiv}
A.~Arrasmith, M.~Cerezo, P.~Czarnik, L.~Cincio, and P.~J. Coles, ``Effect of
  barren plateaus on gradient-free optimization,'' \emph{arXiv:2011.12245},
  2020.

\bibitem{Grimsley2019NatComm}
H.~R. Grimsley, S.~E. Economou, E.~Barnes, and N.~J. Mayhall, ``An adaptive
  variational algorithm for exact molecular simulations on a quantum
  computer,'' \emph{Nat. Commun.}, vol.~10, no.~1, pp. 1--9, 2019.

\bibitem{Sim2019AQT}
\BIBentryALTinterwordspacing
S.~Sim, P.~D. Johnson, and A.~Aspuru‐Guzik, ``Expressibility and {E}ntangling
  {C}apability of {P}arameterized {Q}uantum {C}ircuits for {H}ybrid
  {Q}uantum‐{C}lassical {A}lgorithms,'' \emph{Adv. Quantum Technol.}, vol.~2,
  no.~12, p. 1900070, Oct. 2019. [Online]. Available:
  \url{http://dx.doi.org/10.1002/qute.201900070}
\BIBentrySTDinterwordspacing

\bibitem{Funcke2021Quantum}
\BIBentryALTinterwordspacing
L.~Funcke, T.~Hartung, K.~Jansen, S.~K{\"{u}}hn, and P.~Stornati, ``Dimensional
  {E}xpressivity {A}nalysis of {P}arametric {Q}uantum {C}ircuits,''
  \emph{{Quantum}}, vol.~5, p. 422, Mar. 2021. [Online]. Available:
  \url{https://doi.org/10.22331/q-2021-03-29-422}
\BIBentrySTDinterwordspacing

\bibitem{Li2017PRL}
\BIBentryALTinterwordspacing
J.~Li, X.~Yang, X.~Peng, and C.-P. Sun, ``Hybrid {Q}uantum-{C}lassical
  {A}pproach to {Q}uantum {O}ptimal {C}ontrol,'' \emph{Phys. Rev. Lett.}, vol.
  118, p. 150503, Apr. 2017. [Online]. Available:
  \url{https://link.aps.org/doi/10.1103/PhysRevLett.118.150503}
\BIBentrySTDinterwordspacing

\bibitem{Mitarai2018PRA}
\BIBentryALTinterwordspacing
K.~Mitarai, M.~Negoro, M.~Kitagawa, and K.~Fujii, ``Quantum circuit learning,''
  \emph{Phys. Rev. A}, vol.~98, p. 032309, Sep. 2018. [Online]. Available:
  \url{https://link.aps.org/doi/10.1103/PhysRevA.98.032309}
\BIBentrySTDinterwordspacing

\bibitem{Sweke2020Quantum}
R.~Sweke~\textit{et al}, ``Stochastic gradient descent for hybrid
  quantum-classical optimization,'' \emph{Quantum}, vol.~4, p. 314, 2020.

\bibitem{Harrow2021PRL}
\BIBentryALTinterwordspacing
A.~W. Harrow and J.~C. Napp, ``Low-depth gradient measurements can improve
  convergence in variational hybrid quantum-classical algorithms,'' \emph{Phys.
  Rev. Lett.}, vol. 126, p. 140502, Apr. 2021. [Online]. Available:
  \url{https://link.aps.org/doi/10.1103/PhysRevLett.126.140502}
\BIBentrySTDinterwordspacing

\bibitem{Skolik2021QMI}
A.~Skolik, J.~R. McClean, M.~Mohseni, P.~van~der Smagt, and M.~Leib,
  ``Layerwise learning for quantum neural networks,'' \emph{Quantum Mach.
  Intell.}, vol.~3, no.~1, pp. 1--11, 2021.

\bibitem{Nakanishi2020PRR}
\BIBentryALTinterwordspacing
K.~M. Nakanishi, K.~Fujii, and S.~Todo, ``Sequential minimal optimization for
  quantum-classical hybrid algorithms,'' \emph{Phys. Rev. Research}, vol.~2, p.
  043158, Oct. 2020. [Online]. Available:
  \url{https://link.aps.org/doi/10.1103/PhysRevResearch.2.043158}
\BIBentrySTDinterwordspacing

\bibitem{Ostaszewski2021Quantum}
\BIBentryALTinterwordspacing
M.~Ostaszewski, E.~Grant, and M.~Benedetti, ``Structure optimization for
  parameterized quantum circuits,'' \emph{{Quantum}}, vol.~5, p. 391, Jan.
  2021. [Online]. Available: \url{https://doi.org/10.22331/q-2021-01-28-391}
\BIBentrySTDinterwordspacing

\bibitem{Parrish2019arXiv}
R.~M. Parrish, J.~T. Iosue, A.~Ozaeta, and P.~L. McMahon, ``A {J}acobi
  {D}iagonalization and {A}nderson {A}cceleration {A}lgorithm for {V}ariational
  {Q}uantum {A}lgorithm {P}arameter {O}ptimization,'' \emph{arXiv:1904.03206},
  2019.

\bibitem{WROKS2021}
H.~C. Watanabe, R.~Raymond, Y.-Y. Ohnishi, E.~Kaminishi, and M.~Sugawara,
  ``Optimizing parameterized quantum circuits with free-axis selection,'' in
  \emph{2021 IEEE International Conference on Quantum Computing and Engineering
  (QCE)}, 2021, pp. 100--111.

\bibitem{WROKSYW2022}
\BIBentryALTinterwordspacing
K.~Wada, R.~Raymond, Y.-y. Ohnishi, E.~Kaminishi, M.~Sugawara, N.~Yamamoto, and
  H.~C. Watanabe, ``Simulating time evolution with fully optimized single-qubit
  gates on parametrized quantum circuits,'' \emph{Phys. Rev. A}, vol. 105, p.
  062421, Jun 2022. [Online]. Available:
  \url{https://link.aps.org/doi/10.1103/PhysRevA.105.062421}
\BIBentrySTDinterwordspacing

\bibitem{WRSW2022}
\BIBentryALTinterwordspacing
K.~Wada, R.~Raymond, Y.~Sato, and H.~C. Watanabe, ``Full optimization of a
  single-qubit gate on the generalized sequential quantum optimizer,'' 2022.
  [Online]. Available: \url{https://arxiv.org/abs/2209.08535}
\BIBentrySTDinterwordspacing

\bibitem{nielsen_chuang_2010}
M.~A. Nielsen and I.~L. Chuang, \emph{Quantum Computation and Quantum
  Information: 10th Anniversary Edition}.\hskip 1em plus 0.5em minus
  0.4em\relax Cambridge University Press, 2010.

\bibitem{Kitaev:2002:CQC:863284}
A.~Y. Kitaev, A.~H. Shen, and M.~N. Vyalyi, \emph{Classical and Quantum
  Computation}.\hskip 1em plus 0.5em minus 0.4em\relax Boston, MA, USA:
  American Mathematical Society, 2002.

\bibitem{Kempe:2006:CLH:1122722.1122821}
\BIBentryALTinterwordspacing
J.~Kempe, A.~Kitaev, and O.~Regev, ``The {C}omplexity of the {L}ocal
  {H}amiltonian {P}roblem,'' \emph{SIAM J. Comput.}, vol.~35, no.~5, pp.
  1070--1097, May 2006. [Online]. Available:
  \url{http://dx.doi.org/10.1137/S0097539704445226}
\BIBentrySTDinterwordspacing

\bibitem{Oliveira:2008:CQS:2016985.2016987}
\BIBentryALTinterwordspacing
R.~Oliveira and B.~M. Terhal, ``The complexity of quantum spin systems on a
  two-dimensional square lattice,'' \emph{Quantum Info. Comput.}, vol.~8,
  no.~10, pp. 900--924, Nov. 2008. [Online]. Available:
  \url{http://dl.acm.org/citation.cfm?id=2016985.2016987}
\BIBentrySTDinterwordspacing

\bibitem{Hadfield2020arXiv}
\BIBentryALTinterwordspacing
C.~Hadfield, S.~Bravyi, R.~Raymond, and A.~Mezzacapo, ``Measurements of quantum
  hamiltonians with locally-biased classical shadows,'' \emph{Communications in
  Mathematical Physics}, vol. 391, no.~3, pp. 951--967, May 2022. [Online].
  Available: \url{https://doi.org/10.1007/s00220-022-04343-8}
\BIBentrySTDinterwordspacing

\bibitem{McKay2017}
\BIBentryALTinterwordspacing
D.~C. McKay, C.~J. Wood, S.~Sheldon, J.~M. Chow, and J.~M. Gambetta,
  ``Efficient $z$ gates for quantum computing,'' \emph{Phys. Rev. A}, vol.~96,
  p. 022330, Aug 2017. [Online]. Available:
  \url{https://link.aps.org/doi/10.1103/PhysRevA.96.022330}
\BIBentrySTDinterwordspacing

\bibitem{Qiskit}
H.~Abraham~\textit{et al}, ``Qiskit: {A}n {O}pen-source {F}ramework for
  {Q}uantum {C}omputing,'' 2019.

\bibitem{PhysRevA.71.032313}
\BIBentryALTinterwordspacing
K.~\ifmmode~\dot{Z}\else \.{Z}\fi{}yczkowski and H.-J. Sommers, ``Average
  fidelity between random quantum states,'' \emph{Phys. Rev. A}, vol.~71, p.
  032313, Mar. 2005. [Online]. Available:
  \url{https://link.aps.org/doi/10.1103/PhysRevA.71.032313}
\BIBentrySTDinterwordspacing

\bibitem{fuller2021approximate}
B.~Fuller, C.~Hadfield, J.~R. Glick, T.~Imamichi, T.~Itoko, R.~J. Thompson,
  Y.~Jiao, M.~M. Kagele, A.~W. Blom-Schieber, R.~Raymond \emph{et~al.},
  ``Approximate solutions of combinatorial problems via quantum relaxations,''
  \emph{arXiv preprint arXiv:2111.03167}, 2021.

\bibitem{GW1995}
\BIBentryALTinterwordspacing
M.~X. Goemans and D.~P. Williamson, ``Improved approximation algorithms for
  maximum cut and satisfiability problems using semidefinite programming,''
  \emph{J. ACM}, vol.~42, no.~6, p. 1115–1145, nov 1995. [Online]. Available:
  \url{https://doi.org/10.1145/227683.227684}
\BIBentrySTDinterwordspacing

\bibitem{Moll2018}
N.~Moll, P.~Barkoutsos, L.~S. Bishop, J.~M. Chow, A.~Cross, D.~J. Egger,
  S.~Filipp, A.~Fuhrer, J.~M. Gambetta, M.~Ganzhorn, A.~Kandala, A.~Mezzacapo,
  P.~Müller, W.~Riess, G.~Salis, J.~Smolin, I.~Tavernelli, and K.~Temme,
  ``Quantum optimization using variational algorithms on near-term quantum
  devices,'' \emph{Quantum Science and Technology}, vol.~3, no.~3, p. 030503,
  jun 2018.

\bibitem{gao2021applications}
Q.~Gao~\textit{et al}, ``Applications of quantum computing for investigations
  of electronic transitions in phenylsulfonyl-carbazole {TADF} emitters,''
  \emph{npj Comput. Mater.}, vol.~7, no.~1, pp. 1--9, 2021.

\bibitem{govia2021freedom}
L.~C.~G. Govia, C.~Poole, M.~Saffman, and H.~K. Krovi, ``Freedom of mixer
  rotation-axis improves performance in the quantum approximate optimization
  algorithm,'' 2021.

\bibitem{Candes2009}
\BIBentryALTinterwordspacing
E.~J. Cand{\`e}s and B.~Recht, ``Exact {M}atrix {C}ompletion via {C}onvex
  {O}ptimization,'' \emph{Found. Comput. Math.}, vol.~9, no.~6, p. 717, Apr
  2009. [Online]. Available: \url{https://doi.org/10.1007/s10208-009-9045-5}
\BIBentrySTDinterwordspacing

\bibitem{CandesTao2010}
E.~J. Cand{\`e}s and T.~Tao, ``The {P}ower of {C}onvex {R}elaxation:
  {N}ear-{O}ptimal {M}atrix {C}ompletion,'' \emph{IEEE Trans. Inf. Theory},
  vol.~56, no.~5, pp. 2053--2080, 2010.

\end{thebibliography}
